\documentclass[a4paper,11pt]{article}
\usepackage{jheppub} 
\usepackage[T1]{fontenc} 
\usepackage{array}
\usepackage{subfig}
\newcolumntype{P}[1]{>{\centering\arraybackslash}m{#1}}

\title{\boldmath Measurement of energy flow, cross section \\ and average inelasticity of forward neutrons \\ produced in $\sqrt{s} = 13$~TeV proton-proton collisions \\ with the LHCf Arm2 detector}

\collaborationImg{\includegraphics[width=3.5cm]{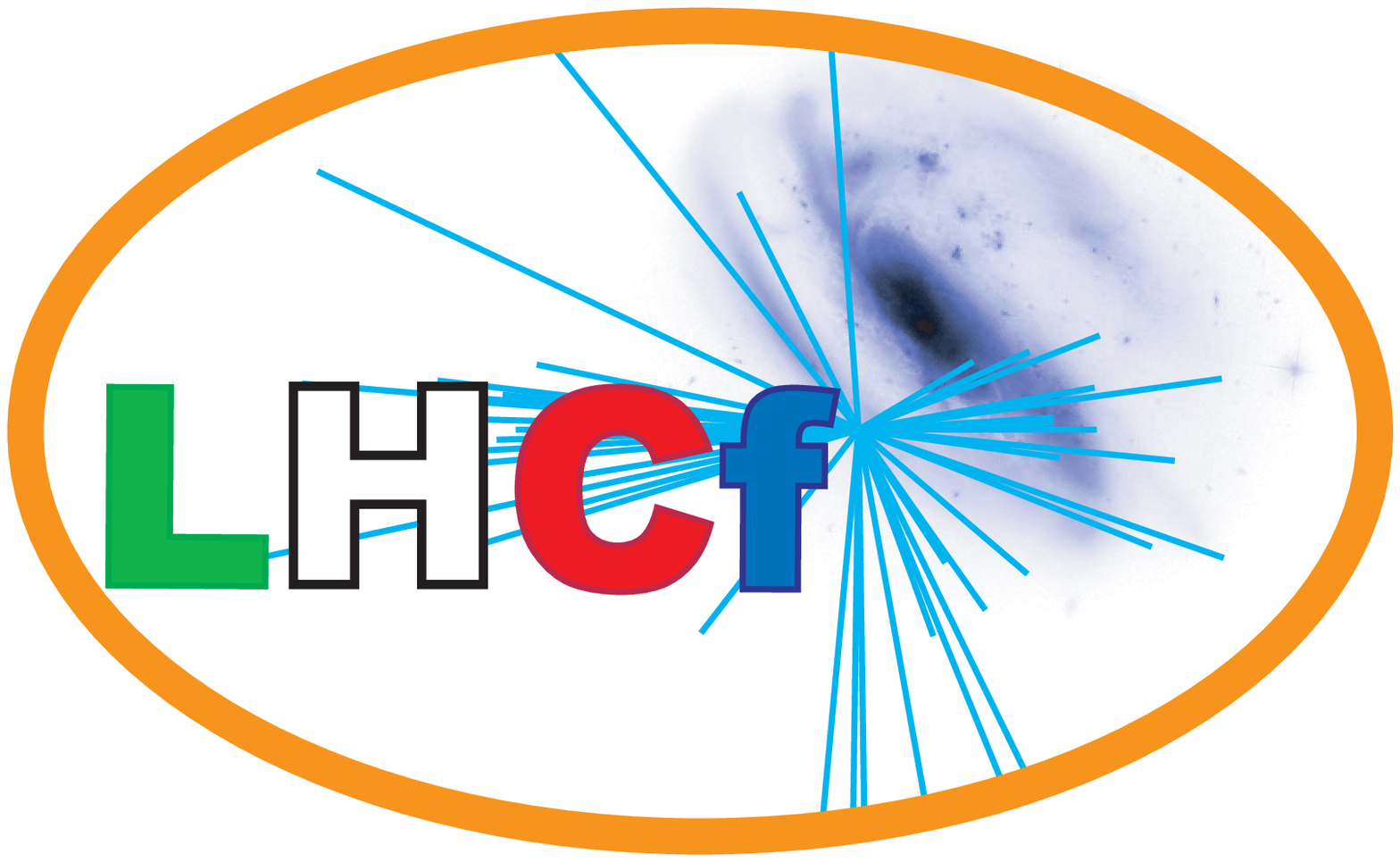} \newline \Large\bfseries\sffamily\raggedright{The LHCf collaboration} \afterCollaborationSpace}

\author[a,b]{O.~Adriani,}
\author[a,b,1]{E.~Berti,\note{Corresponding author.}}
\emailAdd{eugenio.berti@fi.infn.it}
\author[a]{L.~Bonechi,}
\author[a,b]{M.~Bongi,}
\author[a,b]{R.~D'Alessandro,}
\author[a]{S.~Detti,}
\author[c]{M.~Haguenauer,}
\author[d,e]{Y.~Itow,}
\author[f]{K.~Kasahara,}
\author[d]{H.~Menjo,}
\author[d]{Y.~Muraki,}
\author[d]{K.~Ohashi,}
\author[a]{P.~Papini,}
\author[a,g]{S.~Ricciarini,}
\author[h]{T.~Sako,}
\author[i]{N.~Sakurai,}
\author[d]{K.~Sato,}
\author[j]{T.~Tamura,}
\author[a,b]{A.~Tiberio,}
\author[k]{S.~Torii,}
\author[l,m,n]{A.~Tricomi,}
\author[o]{W.C.~Turner,}
\author[d]{M.~Ueno}

\affiliation[a]{INFN Section of Florence, Florence, Italy}
\affiliation[b]{University of Florence, Florence, Italy}
\affiliation[c]{Ecole-Polytechnique, Palaiseau, France}
\affiliation[d]{Institute for Space-Earth Environmental Research, Nagoya~University, Nagoya, Japan}
\affiliation[e]{Kobayashi-Maskawa Institute for the Origin of Particles and the Universe, Nagoya~University, Nagoya, Japan}
\affiliation[f]{Faculty of System Engineering, Shibaura Institute of Technology, Tokyo, Japan}
\affiliation[g]{IFAC-CNR, Florence, Italy}
\affiliation[h]{Institute for Cosmic Ray Research, University of Tokyo, Chiba, Japan}
\affiliation[i]{Tokushima University, Tokushima, Japan}
\affiliation[j]{Kanagawa University, Kanagawa, Japan}
\affiliation[k]{RISE, Waseda University, Shinjuku, Tokyo, Japan}
\affiliation[l]{INFN Section of Catania, Italy}
\affiliation[m]{University of Catania, Catania, Italy}
\affiliation[n]{CSFNSM, Catania, Italy}
\affiliation[o]{LBNL, Berkeley, California, USA}

\abstract{
In this paper, we report the measurement of the energy flow, the cross section and the average inelasticity of forward neutrons (+ antineutrons) produced in $\sqrt{s} = 13$~TeV proton-proton collisions. These quantities are obtained from the inclusive differential production cross section, measured using the LHCf Arm2 detector at the CERN Large Hadron Collider. The measurements are performed in six pseudorapidity regions: three of them ($\eta > 10.75$, $8.99 < \eta < 9.21$ and $8.80 < \eta < 8.99$), albeit with smaller acceptance and larger uncertainties, were already published in a previous work, whereas the remaining three ($10.06 < \eta < 10.75$, $9.65 < \eta < 10.06$ and $8.65 < \eta < 8.80$) are presented here for the first time. The analysis was carried out using a data set acquired in June 2015 with a corresponding integrated luminosity of $\mathrm{0.194~nb^{-1}}$. Comparing the experimental measurements with the expectations of several hadronic interaction models used to simulate cosmic ray air showers, none of these generators resulted to have a satisfactory agreement in all the phase space selected for the analysis. The inclusive differential production cross section for $\eta > 10.75$ is not reproduced by any model, whereas the results still indicate a significant but less serious deviation at lower pseudorapidities. Depending on the pseudorapidity region, the generators showing the best overall agreement with data are either SIBYLL 2.3 or EPOS-LHC. Furthermore, apart from the most forward region, the derived energy flow and cross section distributions are best reproduced by EPOS-LHC. Finally, even if none of the models describe the elasticity distribution in a satisfactory way, the extracted average inelasticity is consistent with the QGSJET II-04 value, while most of the other generators give values that lie just outside the experimental uncertainties.
}


\notoc

\begin{document} 
\maketitle
\flushbottom

\section{Introduction}

In the last decade, gigantic cosmic ray air shower experiments, like the Pierre Auger Observatory \cite{ref:PAO} and the Telescope Array \cite{ref:TA}, have contributed to fundamental progress in the knowledge of Ultra-High Energy Cosmic Rays (UHECRs), \textit{i.e.} cosmic rays with energies above $10^{18}$~eV \cite{ref:Review}. Thanks to the large statistics and high quality results obtained by these experiments, various theories have been suggested to explain the mechanisms responsible for acceleration and propagation of UHECRs. However, despite this significant progress, the current measurements leave several open questions on the nature of UHECRs, mainly relative to their astrophysical sources, acceleration mechanisms and mass composition at high energies. 

In gigantic cosmic ray air shower experiments, the properties of the primary cosmic ray particles are indirectly reconstructed from the detection of Extensive Air Showers (EASs), \textit{i.e.} the showers of secondary particles that primary particles form when they interact in the atmosphere. In particular, mass composition is not a direct information, but can be derived from the measurement of the depth at which the number of particles in the shower reaches its maximum, or from the simultaneous measurements of the number of muons and electrons at the ground level. In order to extract this information, the experimental results must be compared to simulations performed under different assumptions on the relative mass abundance of the primary cosmic rays. Nevertheless, due to the lack of high energy calibration data, the predictions of hadronic interaction models used to simulate the EASs are subject to large systematic errors, which in turn constitute the major source of uncertainty on mass composition measurements \cite{ref:Unger}. To reduce this systematic contribution, accurate information on several parameters used to model EAS development are needed: inelastic cross section, multiplicity of secondaries and forward energy distributions, from which one can derive the average inelasticity \cite{ref:Matthews}\cite{ref:Ulrich}. This information can be extracted from measurements carried out at hadron colliders and the CERN Large Hadron Collider (LHC) \cite{ref:LHC} is currently the best candidate to accomplish this task. During LHC Run II, the machine operated p-p collisions at $\sqrt{s} = 13$ TeV, which is the highest energy ever achieved at a collider and in the range relevant for UHECRs (equivalent to the collision of a $\mathrm{0.9 \times 10^{16}~eV}$ proton with a nucleon of a nucleus in the atmosphere). At LHC, inelastic cross section and multiplicity of secondaries are mainly accessible to the four central detectors \cite{ref:ATLAS, ref:CMS, ref:ALICE, ref:LHCB} and roman pot detectors like TOTEM \cite{ref:TOTEM} and ATLAS ALFA \cite{ref:ALFA}, while forward energy distributions and average inelasticity can be measured only with dedicated forward detectors. Some of the results obtained during LHC Run I from the first group of experiments (\textit{e.g.} \cite{ref:ATLAS_measurement, ref:CMS_measurement, ref:ALICE_measurement, ref:TOTEM_measurement}) were used to tune the so-called post-LHC models, like QGSJET II-04 \cite{ref:QGSJET}, EPOS-LHC \cite{ref:EPOS} and SIBYLL 2.3 \cite{ref:SIBYLL}. However, the level of agreement among these models is still far from satisfactory \cite{ref:PAO_model}, thus highlighting the importance of the second group of experiments. Even if several LHC experiments have forward subdetectors (see \cite{ref:Forward} for a complete list of forward detectors operating during Run II), LHCf \cite{ref:LHCf_TDR} is the only one that has been specifically designed to accurately measure the distributions of very forward neutral particles produced in proton-proton collisions.

Because of the key role that forward baryons have in the development of the air showers and in the abundance of the muonic component \cite{ref:Pierog}, a large activity of the LHCf collaboration is dedicated to neutron measurements. A first result relative to the inclusive differential production cross section of very forward neutrons+antineutrons (hereafter simply called \textit{neutrons}) produced in p-p collisions at $\sqrt{s} = 13$~TeV was already published \cite{ref:LHCf_Neutron}. Here this analysis is extended from three to six  pseudorapidity regions, in order to have enough data points to derive three important quantities that are directly connected to EAS development: neutron energy flow, cross section and average inelasticity. It is worth mentioning that these are the first measurements of this type at such a high energy and that their direct comparison to generator expectations gives an important indication on the validity of these generators for air shower simulation.

\section{The LHCf experiment}

The LHCf experiment consists of two detectors \cite{ref:LHCf_detector}, Arm1 and Arm2, placed in two regions on the opposite sides of LHC Interaction Point 1 (IP1). These regions, called Target Neutral Absorber (TAN), are located at a distance of 141.05~m from IP1, after the dipole magnets that bend the two proton beams. In this position, the two detectors are capable of detecting the neutral particles that are produced in hadronic collisions with a pseudorapidity $\eta > 8.4$.

The measurements reported in this paper are obtained using the LHCf Arm2 detector only, because, due to the different geometric acceptance of the two arms, it offers a better coverage of the pseudorapidity regions where the core of the energy flow is expected\footnote{However, for the three pseudorapidity regions already published in \cite{ref:LHCf_Neutron}, the consistency between the preliminary results obtained with the Arm1 and the Arm2 detectors was checked and confirmed.
}. The Arm2 detector consists of two calorimetric towers with different transverse sizes: the \textit{small tower} (25~mm $\times$ 25~mm) and the \textit{large tower} (32~mm $\times$ 32~mm). Each tower is made out of 16 $\mathrm{Gd_{2}SiO_{5}}$ (GSO) scintillator layers (1~mm thick), interleaved with 22 tungsten (W) plates (7~mm thick), for a total length of about 21~cm, which is equivalent to 44~$\mathrm{X_{0}}$ or 1.6~$\mathrm{\lambda _{I}}$. In addition, 4 xy imaging layers, made out of silicon microstrip detectors with a read-out pitch of $\mathrm{160~\mu m}$, are inserted at different depths. In this way, for all particles hitting the detector, it is possible to reconstruct the incident energy (from the scintillator layers)  and the transverse position (from the imaging layers). Concerning the reconstruction performances for incident hadrons at the two different energies of $\mathrm{1~TeV}$ and $\mathrm{6.5~TeV}$, the detection efficiency increases from $52\%$ to $72\%$, the energy resolution worsens from $28\%$ to $38\%$, while the position resolution improves from $\mathrm{300~\mu m}$ to $\mathrm{100~\mu m}$ \cite{ref:LHCf_performances}.

Thanks to the detector design, photons and hadrons are easily separated using the different features of electromagnetic and hadronic showers. However, being unable to distinguish among hadrons, the experimental measurement contains not only neutrons and antineutrons, but also $\mathrm{\Lambda^{0}}$, $\mathrm{K^{0}_{L}}$ and other neutral and charged hadrons\footnote{
These charged hadrons, mainly $\mathrm{\pi^{\pm}}$ and protons, are either the products of the decay of a neutral hadron after the magnets or particles directly produced in the collisions and bended by the magnets.
}. As described later, this residual background is subtracted at the end of the analysis by the usage of Monte Carlo simulations. 

\section{Experimental data set}
\label{sec:exp}

The results discussed in this paper are based on an experimental data set, relative to p-p collisions at $\sqrt{s} = 13$~TeV, which was acquired from 22:32 of June 12th to 1:30 of June 13th 2015 (CEST). This time period corresponds to LHC Fill 3855, a special low luminosity and high $\beta^{*}$ fill specifically provided for LHCf operations. Low luminosity ($\mathrm{3-5\times 10^{28}~cm^{-2} s^{-1}}$) was required in order to keep a small average number of collisions per bunch crossing ($0.007-0.012$). In this way, given the $15\%$ acceptance of the calorimeter for inelastic collisions, event pile-up at the detector level remains below a negligible $1\%$. High $\beta^{*}$ ($\mathrm{19~m}$) was required in order to keep the protons composing the beam almost parallel at the interaction point. In this way, the scattering angle (and therefore the pseudorapidity) of the detected secondary particles can be accurately reconstructed. In Fill 3855, 29 bunches were made to collide at IP1 with a downward half crossing angle of 145~$\mathrm{\mu rad}$, a value chosen to increase the detector coverage down to a pseudorapidity of $8.4$. In addition, there were 6 and 2 non-colliding bunches circulating in the clockwise and counter-clockwise beams, respectively. Using the instantaneous luminosity measured from the ATLAS experiment \cite{ref:ATLAS-luminosity} and taking into account LHCf data acquisition live time, the recorded integrated luminosity was estimated to be $\mathrm{0.194~nb^{-1}}$. 

\section{Simulation data set}
\label{sec:mc}

Monte Carlo simulations with the same experimental configuration used for LHC data taking are necessary for four different purposes: estimation of correction factors and systematic uncertainties, validation of the whole analysis procedure, energy spectra unfolding, and data-model comparison. A description of all the simulation data sets, detailing how they were generated, which models were used and which effects were considered, can be found in \cite{ref:LHCf_Neutron}. Here it is important to remind that, depending on the purpose, a given sample is generated taking into account one or more of the following steps: hadronic collision, transport of secondary products from IP1 to TAN, and detector interactions. The Cosmos 7.633 \cite{ref:COSMOS} and EPICS 9.15 \cite{ref:EPICS} libraries were used for all these simulation steps, except for the samples needed for the final data-model comparison. In this last case, QGSJET II-04, EPOS-LHC, SIBYLL 2.3 and DPMJET 3.06 \cite{ref:DPMJET} data sets were simulated using CRMC \cite{ref:CRMC}, an interface tool for event generators, whereas the PYTHIA 8.212 \cite{ref:PYTHIA} data set was simulated using its own dedicated interface.

\section{Analysis}

The analysis procedure presented in this section is divided in four main steps. First, after defining the event selection criteria, the events in the sample are reconstructed in order to derive the detector level energy distributions for each pseudorapidity region (section \ref{sec:reconstruction}). Second, the correction factors that must be applied to the spectra before and after unfolding are estimated (section \ref{sec:corrections}). Third, the effect of migration between energy bins is unfolded using a deconvolution procedure that takes into account the detector response (section \ref{sec:unfolding}). Fourth, all necessary systematic uncertainties are evaluated and associated to the unfolded spectra (section \ref{sec:systematics}). Finally, these distributions are used to derive the experimental measurements, which are then compared to the various generator predictions. These results are presented and discussed in the last two sections, relative to the inclusive differential production cross section (section \ref{sec:results}) and to several quantities connected to air shower development (section \ref{sec:discussion}).

\subsection{Event reconstruction}
\label{sec:reconstruction}

\begin{figure*}[tbp]
 \centering
 \includegraphics[width=0.5\textwidth]{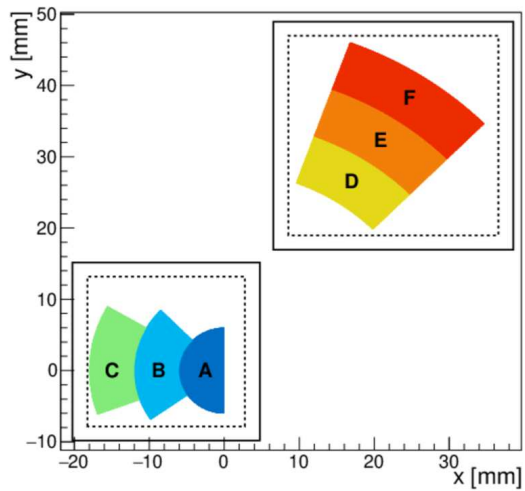}
 \caption{ 
  Definition of the six pseudorapidity regions used in the analysis. Bottom left and upper right squares respectively correspond to the small and the large tower of the Arm2 detector as seen from IP1. The origin of the reference frame is centered in the beam center projection on the detector plane during LHC Fill 3855. All analysis regions are chosen within a fiducial area (dashed line), which is 2~mm inside the tower edges (solid line).   
  }
 \label{fig:regions}
\end{figure*}

\enlargethispage{+2\baselineskip}
Apart from small refinements, event reconstruction algorithms and event selection criteria are similar to the ones described in \cite{ref:LHCf_Neutron}. The incident energy is reconstructed from the total energy deposit in the calorimeter, properly weighting the release in each scintillator layer for the energy lost in the corresponding front tungsten layer and applying position dependent correction factors for shower lateral leakage and light collection efficiency. The transverse position is determined by fitting the shower lateral profile reconstructed in the imaging layers, always looking for a \textit{singlehit} event, \textit{i.e.} a single peak in each tower. In case of a \textit{multihit} event, \textit{i.e.} more simultaneous hits in the same tower, the transverse position is obviously misreconstructed, an effect that is later corrected for in the analysis. An event is accepted if it satisfies an offline selection that mimics the hardware trigger, but with a slightly higher threshold respect to the one used by the discriminators during data taking. 
\begin{table}[tbp]
  \begin{center}
    \begin{tabular}{|c|c|c|c|c|c|c|}
      \hline
      & \textbf{Region A} & \textbf{Region B} & \textbf{Region C} & \textbf{Region D} & \textbf{Region E} & \textbf{Region F} \\
      \hline
	  $\eta$
	  & 10.75--$\infty$	& 10.06--10.75	 & 9.65--10.06 & 8.99--9.21 & 8.80--8.99 & 8.65--8.80\\
	  $\mathrm{\theta~[\mu rad]}$	
	  & 0--42	& 42--85	& 85--128	& 198--249	& 249--298	& 298--347\\
	  $\phi~[\ensuremath{^\circ}]$	
	  & 90--270 & 135--215 & 150--200 & 45--70 & 45--70 & 45--70\\
      \hline
    \end{tabular}
    \caption{Definition of the six pseudorapidity regions used in the analysis: the coverage is expressed in terms of pseudorapidity $\eta$, scattering angle $\theta$ and azimuthal angle $\phi$.}
    \label{tab:regions}
  \end{center}
\end{table}
This offline selection requires a raw energy deposit above $\mathrm{850~MeV}$ in at least three consecutive scintillator layers, ensuring a good selection efficiency for hadrons with incident energies above $\mathrm{500~GeV}$. Particle IDentification (PID) exploits the different development of electromagnetic and hadronic showers in the calorimeter to distinguish between photon and hadron candidates. This discrimination is based on the variable $L_{2D} = L_{90\%} - 0.25 \times L_{20\%}$, where $L_{X\%}$ represents the longitudinal depth at which the fraction of the energy deposit with respect to the total release in the calorimeter is $X\%$. Finally, after taking into account the beam center projection on the detector plane, the event is selected if it is within one of the six pseudorapidity regions defined for this analysis. 
\enlargethispage{+1\baselineskip}
Three of these pseudorapidity intervals ($\eta > 10.75$, $8.99 < \eta < 9.21$ and $8.80 < \eta < 8.99$) were already considered in \cite{ref:LHCf_Neutron}, whereas the remaining three ($10.06 < \eta < 10.75$, $9.65 < \eta < 10.06$ and $8.65 < \eta < 8.80$) are new additions to this work. The six regions superimposed to the detector area are shown in figure \ref{fig:regions} and more details on their definition are given in table \ref{tab:regions}. Due to the slightly different definition of the three regions already included in \cite{ref:LHCf_Neutron}, and to the refined reconstruction algorithm and selection criteria, the correction factors and the systematic uncertainties discussed here may differ of a few percent from the values reported there.

\subsection{Correction factors}
\label{sec:corrections}

\begin{table}[tbp]
  \begin{center}
    \begin{tabular}{|c|c|c|c|}
       	\hline
		\textbf{Correction} & \textbf{Region A} & \textbf{Region B} & \textbf{Region C} \\
    	\hline
		Beam background	 	 &	 (-3\%, -1\%)  	&	 (-3\%, -1\%)   & (-12\%, -1\%)\\
		PID selection  	 	 &	 (-14\%, +7\%) 	&	 (-9\%, +8\%)   & (-9\%, +11\%)\\
		Multihit events	 	 &	 (-10\%, +9\%) 	&	 (-10\%, +7\%)  & (-16\%, +3\%)\\
		Fake events	  	 	 &  (-8\%, -1\%) 	&	 (-9\%, -1\%)  	& (-8\%, -1\%)\\
    	\hline
		Missed events  	  	 &  (+38\%, +269\%)&    (+37\%, +245\%)& (+38\%, +286\%)\\  
		Hadron contamination &  (-62\%, -6\%)  &  	 (-60\%, -8\%)  & (-59\%, -11\%)\\     
		\hline
    \end{tabular}
    \hspace{5cm}
    \begin{tabular}{c}
    \end{tabular}
    \hspace{5cm}
    \begin{tabular}{|c|c|c|c|}
       	\hline
		\textbf{Correction} & \textbf{Region D} & \textbf{Region E} & \textbf{Region F} \\
    	\hline
		Beam background   	  &	 (-4\%, -1\%)  	&	 (-5\%, -1\%)   & (-4\%, -1\%)\\
		PID selection	  	  &	 (-6\%, +6\%)  	&	 (-3\%, +8\%)   & (-3\%, +10\%)\\
		Multihit events	  	  &	 (-11\%, +3\%) 	&	 (-14\%, +2\%)  & (-20\%, +2\%)\\
		Fake events  	  	  &  (-6\%, -1\%) 	&	 (-6\%, -1\%)   & (-6\%, -1\%)\\
    	\hline
		Missed events 	  	  &  (+35\%, +239\%)&    (+33\%, +222\%)& (+32\%, +229\%)\\  
		Hadron contamination  &  (-54\%, -15\%) &    (-54\%, -14\%) & (-52\%, -18\%)\\     
		\hline
    \end{tabular}
	\caption{The (minimum, maximum) values of the relative correction factors for each contribution discussed here, separately reported for regions A, B, C, D, E and F. The top four and the bottom two corrections are respectively applied before and after unfolding.}
    \label{tab:corrections}
  \end{center}
\end{table}

Correction factors account for the same effects and approximately amount to the same values described in \cite{ref:LHCf_Neutron}. All contributions are summarized in table \ref{tab:corrections} and are shortly discussed in the following. Note that correction factors are divided into two groups, depending on whether they are applied before or after unfolding. The reason for this distinction is that, as explained in section \ref{sec:unfolding}, the final result is obtained making use of energy unfolding based on a single hadron response matrix. Thus, the measured distributions must be corrected before unfolding for all the effects that deviate from the single hadron case. The remaining corrections are applied after unfolding in order to simplify the analysis procedure.

Four correction factors are applied before unfolding. The first correction is due to \textit{beam background}, caused by two different processes: the interaction of primary protons with the residual gas in the beam line and the interaction of secondary particles from collisions with the beam pipe. The former, estimated using non-colliding bunches, leads to a correction of about $-1\%$, whereas the latter, estimated using dedicated simulations, ranges from $-10\%$ to $0\%$, depending on the energy. The second correction accounts for the limited efficiency (hadron identification) and purity (photon contamination) of the \textit{PID selection}. Correction factors, estimated via template fit of the $L_{2D}$ distributions of simulated photons and hadrons to the experimental data, range between $-15\%$ and $+10\%$. The uncertainty of this correction, obtained from the combination of a dedicated confidence interval fit and different template fit strategies, is generally below $20\%$. The third correction is due to \textit{multihit events}, which cannot be treated as a simple background to be subtracted, because this choice would not lead to inclusive measurements. Instead, the corrections applied should restore the ideal case distributions where one is able to reconstruct the $n$ simultaneous hits in the same tower as $n$ independent events. Exploiting two simulation samples generated using QGSJET II-04 and EPOS-LHC, the correction factors, estimated from their average, are in a range between $-20\%$ and $+10\%$, with an uncertainty, given by half of their deviation, mostly below $10\%$. The fourth correction accounts for \textit{fake events}, \textit{i.e.} the events that, due to detector misreconstructions, are incorrectly included in the measured distributions. Correction factors, estimated employing the same simulation samples and the same computation method just described, range from $-10\%$ to $-1\%$, whereas the uncertainty is always smaller than $1\%$.

Two correction factors are applied after unfolding. The first correction accounts for \textit{missed events}, \textit{i.e.} the events that, due to detector misreconstructions, are incorrectly excluded from the measured distributions. Correction factors, estimated employing the same simulation samples and the same computation method used for fake events, range from $+30\%$ to more than $+100\%$, whereas the uncertainty is always smaller than $1\%$. As expected, missed events are the dominant contribution to correction factors because of the limited hadron detection efficiency, which decreases even more at low energies. Two factors contribute to the uncertainty on fake and missed corrections. The first one is the dependence on the simulation model used to generate the collisions, which, as discussed in this section, is negligible respect to the other contributions. The second one is the dependence on the simulation model used to describe the interaction with the detector, which, as discussed in section \ref{sec:systematics}, is not negligible. The last correction is due to \textit{hadron contamination}, \textit{i.e.} the fact that a non-negligible fraction of hadrons reaching the detector are not actually neutrons, but other particles that cannot be distinguished from them. Exploiting five simulation samples generated using all the models discussed in section \ref{sec:mc}, the correction factors, estimated from their average, are in a range between $-60\%$ and $-5\%$, with an uncertainty, given by their maximum deviation, mostly below $20\%$.

\subsection{Spectra unfolding}
\label{sec:unfolding}

Given the limited hadron energy resolution, spectra unfolding must be applied to deconvolute the measurements from the detector response. The iterative Bayesian method \cite{ref:DAgostini}, implemented in the RooUnfold package \cite{ref:RooUnfold}, was used for this purpose. The best estimation for the final unfolded distributions was obtained choosing a flat prior as initial hypothesis and constructing the response matrix from a single neutron flat energy sample simulated using the DPMJET 3.04 model. As discussed in section \ref{sec:systematics}, in order to estimate the uncertainties due to the deconvolution, the unfolding procedure was repeated several times with different priors, response matrices, and test input spectra. As a general convergence criteria, the iterative method was stopped when $\Delta \chi^{2}$, the $\chi^{2}$ variation between the outputs of two consecutive iterations, was below 1. This threshold was chosen as a compromise between convergence requirements and the number of iterations needed to reach that convergence. As an example, in the case of the data points in the final distributions, this convergence criteria corresponds to 51, 35, 21, 13, 11 and 10 iterations for pseudorapidity regions A, B, C, D, E and F, respectively.

\subsection{Systematic uncertainties}
\label{sec:systematics}

\begin{table}[tbp]
  \begin{center}
    \begin{tabular}{|c|c|c|c|}
       	\hline
		\textbf{Systematic} & \textbf{Region A} & \textbf{Region B} & \textbf{Region C} \\
    	\hline
		Energy scale			& 2--133\% &	 1--120\% &  0--94\% \\
		Beam center 			& 0--6\%   &	 0--11\%  &	 0--10\% \\
		Position resolution 	& 0--41\%  & 	 0--31\%  &	 0--28\% \\
		PID selection 	 		& 0--24\%  &	 0--32\%  &	 0--25\% \\
		Multihit events 		& 0--13\%  &	 0--8\%   &	 0--10\%  \\
		\hline
		Unfolding method 	 	& 0--22\%  &	 0--20\%  &	 1--33\%  \\
		Unfolding prior	 		& 3--19\%  &	 2--39\%  &	 3--69\%  \\
		Unfolding algorithm 	& 2--100\% &	 1--100\% &	 4--100\%  \\
		Hadron contamination  	& 3--24\%  &	 3--21\%  &	 5--21\%  \\  
		\hline
    \end{tabular}
    \hspace{5cm}
    \begin{tabular}{c}
    \end{tabular}
    \hspace{5cm}
    \begin{tabular}{|c|c|c|c|}
       	\hline
		\textbf{Systematic} & \textbf{Region D} & \textbf{Region E} & \textbf{Region F} \\
    	\hline
		Energy scale 			& 0--77\%  &	 0--68\%  &  1--68\% \\
		Beam center 			& 0--8\%   &	 0--10\%  &	 0--9\% \\		
		Position resolution 	& 0--9\%   &  	 0--10\%  &	 0--9\% \\
		PID selection 	 		& 1--14\%  &	 1--12\%  &	 1--18\% \\
		Multihit events 		& 0--3\%   &	 0--6\%   &	 0--6\%  \\
		\hline
		Unfolding method 	 	& 0--47\%  & 	 1--48\%  &	 2--33\%  \\
		Unfolding prior 	 	& 3--66\%  &	 0--51\%  &	 1--60\%  \\
		Unfolding algorithm 	& 1--100\% &	 0--100\% &	 0--140\%  \\
		Hadron contamination  	& 5--18\%  &	 5--17\%  &	 6--17\%  \\  	
		\hline	
    \end{tabular}
	\caption{The (minimum, maximum) values of the relative systematic uncertainties for each contribution discussed here, separately reported for regions A, B, C, D, E and F. The top five and the bottom four uncertainties are respectively applied before and after unfolding. Note that the two energy independent systematic uncertainties, relative to \textit{integrated luminosity} ($1.9\%$) and \textit{detection efficiency} ($6\%$), applied after unfolding, are not reported in this table.}
    \label{tab:systematics}
  \end{center}
\end{table}
		
Systematic uncertainties are due to the same effects and approximately amount to the same values described in \cite{ref:LHCf_Neutron}. Considering all the uncertainty sources as independent, the final uncertainty is obtained from their quadrature sum. All contributions are summarized in table \ref{tab:systematics} and are shortly discussed in the following.  Note that systematic uncertainties are divided into two groups, depending on whether they are applied before or after unfolding. The reason for this distinction is that some uncertainties,  which are relative to experimental conditions or detector response, can be estimated only on the measured distributions, whereas other uncertainties, which are energy independent or unfolding related, can be directly applied on the unfolded distributions. Thus, while the latter are immediately added to the final results, the former must be fully propagated through the unfolding procedure. 

Three uncertainties are applied before unfolding. The first uncertainty is relative to the \textit{energy scale}, which is known with an error of $\pm 4.5\%$, determined from the detector calibration at SPS, the LHC data taking conditions, and the observed $\pi^{0}$ mass peak shift. The contribution to the final uncertainty, estimated by shifting the energy scale for the corresponding error and comparing them with the nominal distribution, increases from a value of about $10\%$ at low energy up to more than $100\%$ at high energy, thus representing the dominant contribution to the systematic uncertainty. The second uncertainty is associated to the \textit{beam center} projection on the detector plane, which was determined by fitting the position distribution of high energy hadrons with a two dimensional function, leading to an error of $\pm 0.3$~mm. The corresponding uncertainty, obtained by shifting the beam center for the corresponding error on the two axes and comparing them with the nominal distribution, is generally below $10\%$. The third uncertainty comes from the \textit{position resolution}, which is not taken into account in the deconvolution in order to simplify the unfolding procedure. Exploiting two simulation samples generated with QGSJET II-04 and EPOS-LHC, the corresponding uncertainty, estimated from the ratio between the spectra obtained using true and reconstructed position, is mostly below $10\%$. Note that, for each event, these three uncertainties affect the reconstructed energy (energy scale and position resolution) and/or the reconstructed pseudorapidity (beam center and position resolution).

Three uncertainties are applied after unfolding. The first uncertainty is associated to the \textit{integrated luminosity}, as derived from ATLAS measurements, and has an energy independent value of $1.9\%$. The second uncertainty comes from the \textit{detection efficiency}, which is affected by the error on the hadron-detector interaction cross section at high energies. In order to estimate the impact of this parameter on the experimental measurements, its value was extracted both for the DPMJET 3.04 model in EPICS and for the QGSP\_BERT 4.0 model in GEANT4 \cite{ref:GEANT4}. Then, assuming that the difference on the hadron-detector interaction cross section leads to a similar difference on the detection efficiency, the corresponding uncertainty is estimated from the maximum deviation below $\mathrm{6.5~TeV}$, amounting to an energy independent value of $6\%$. The third uncertainty accounts for three different contributions relative to \textit{spectrum unfolding}: \textit{method} uncertainty, due to the deviation between unfolded spectrum and true distribution, estimated using different generators; \textit{prior} uncertainty, due to the dependence of the unfolded spectrum on the prior chosen as input of the algorithm, estimated using different priors; \textit{algorithm} uncertainty, due to the dependence of the unfolded result on the deconvolution algorithm, estimated using different approaches (Bayesian, SVD \cite{ref:SVD} and TUnfold \cite{ref:TUnfold}). Each of these three uncertainties is mostly below $20\%$, except at very high energies, where they can increase to more than $100\%$, due to the small number of events.  

Finally, additional sources of uncertainties are associated to the \textit{correction factors} discussed in section \ref{sec:corrections}: in this case, each of these contributions belongs to the group of uncertainties applied either before or after unfolding, depending on where the corresponding correction is considered in the analysis.

\section{Results}
\label{sec:results}

\begin{figure*}[tbp]
 \centering
 \includegraphics[width=.333\textwidth]{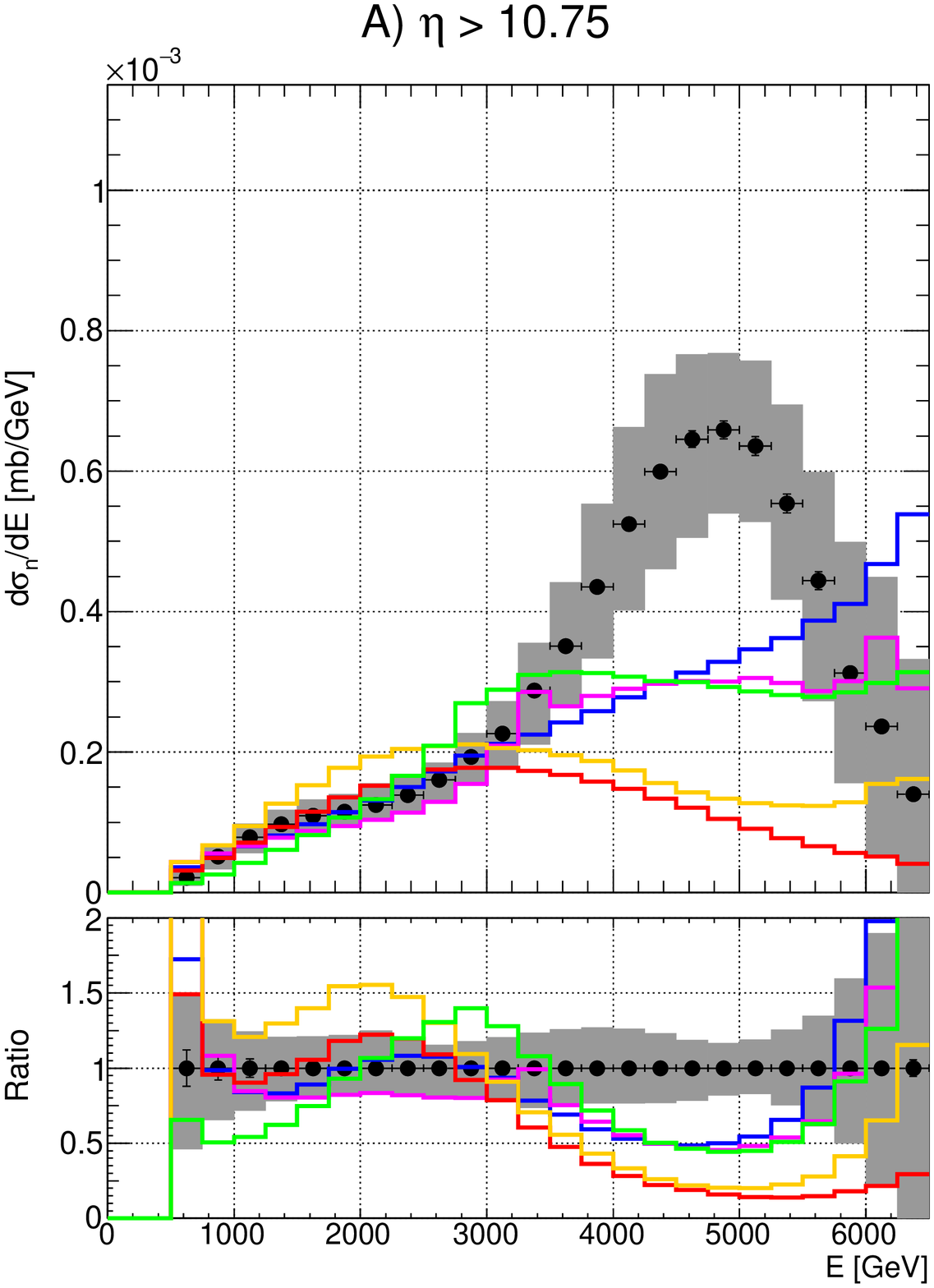}%
 \includegraphics[width=.333\textwidth]{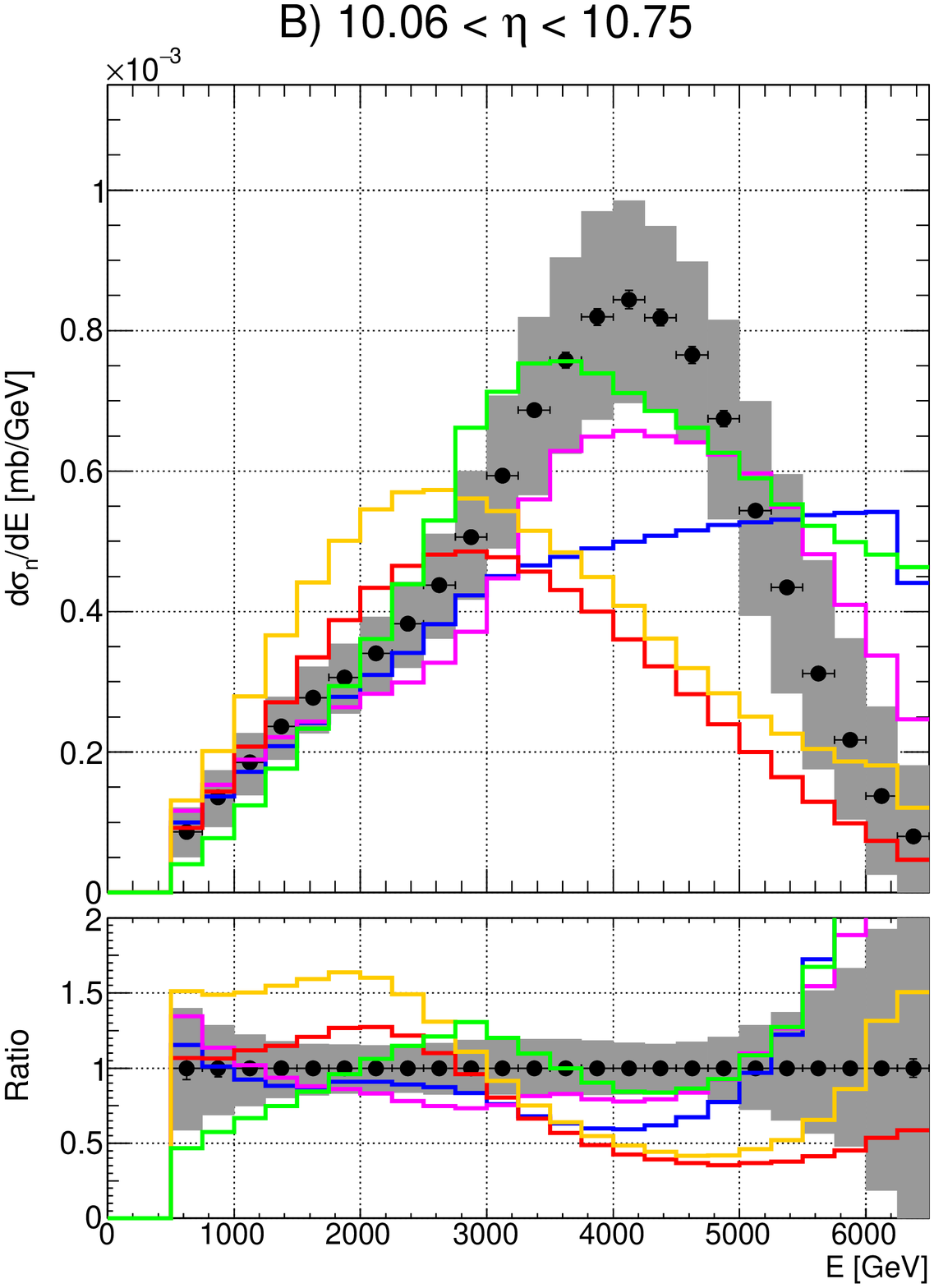}%
 \includegraphics[width=.333\textwidth]{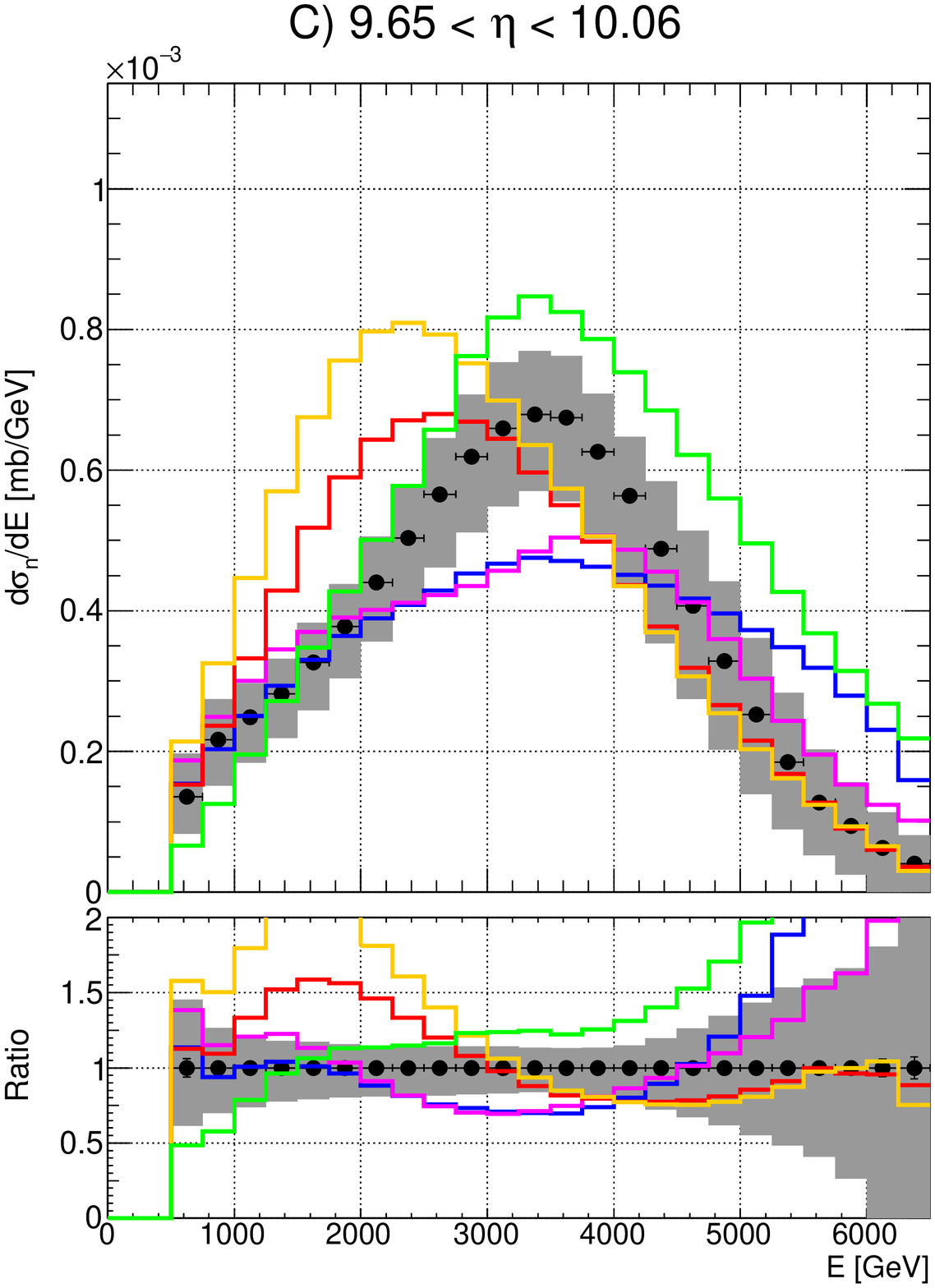}%
  \hspace{0.1mm}
 \includegraphics[width=.333\textwidth]{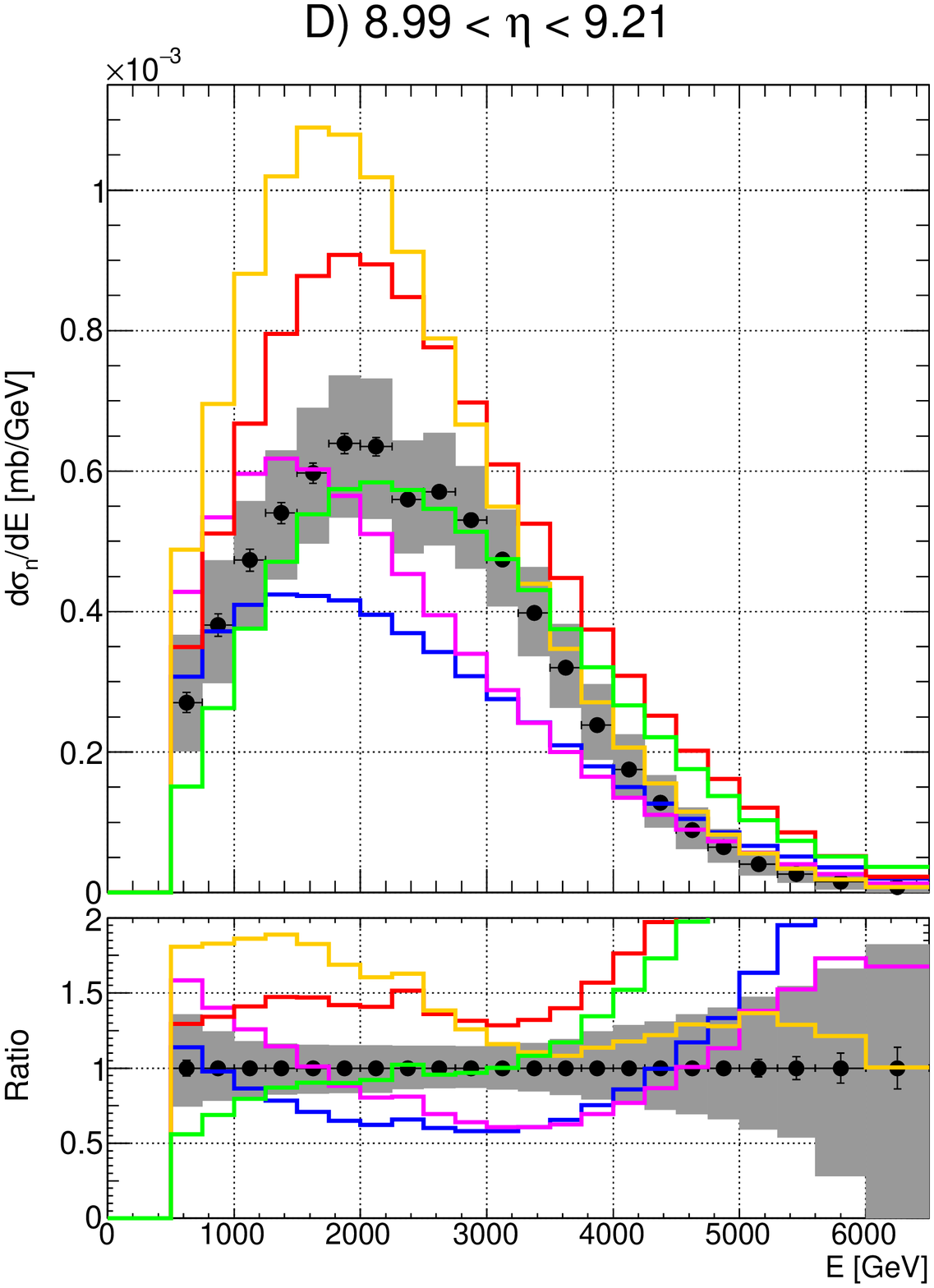}%
 \includegraphics[width=.333\textwidth]{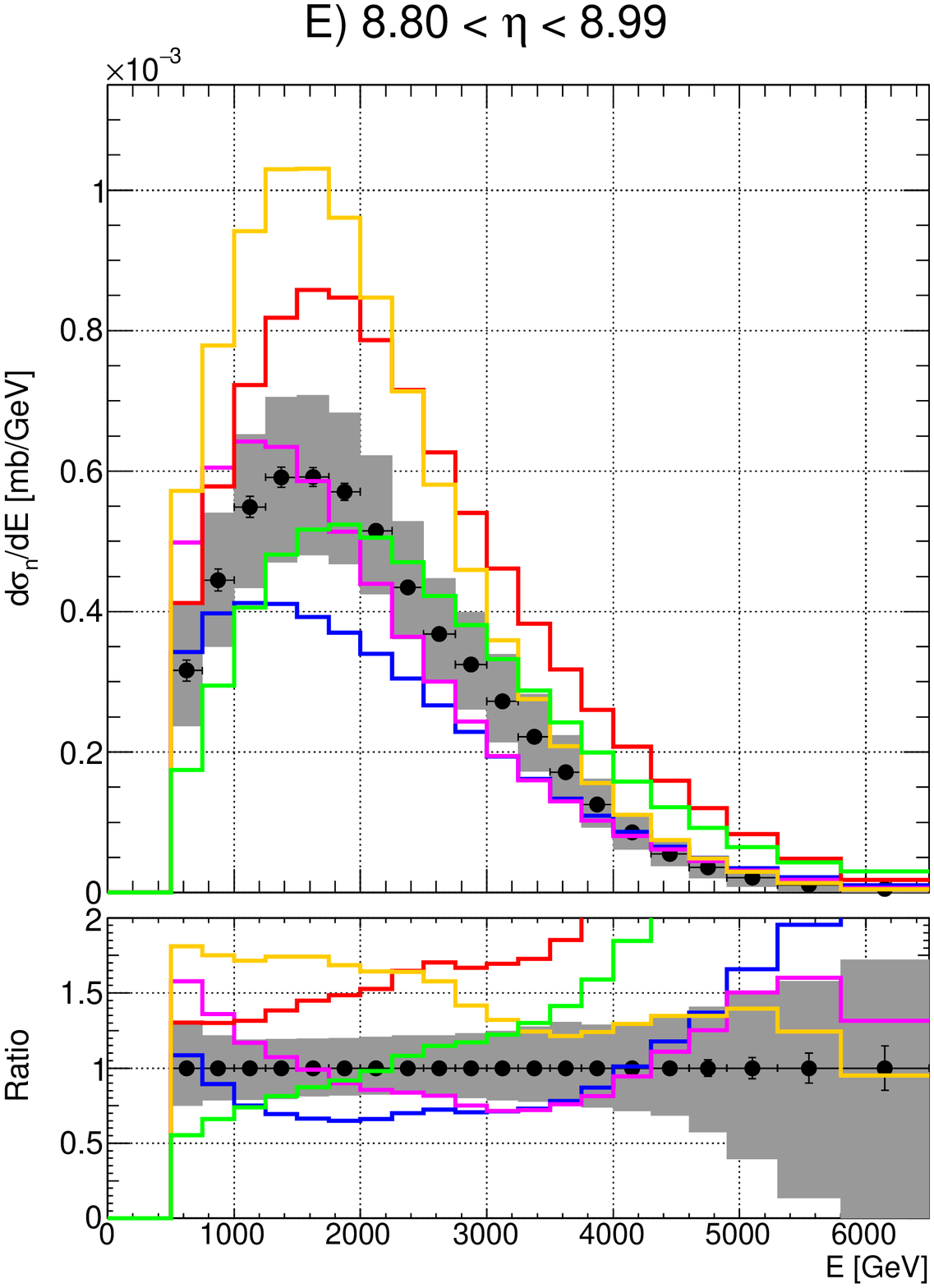}%
 \includegraphics[width=.333\textwidth]{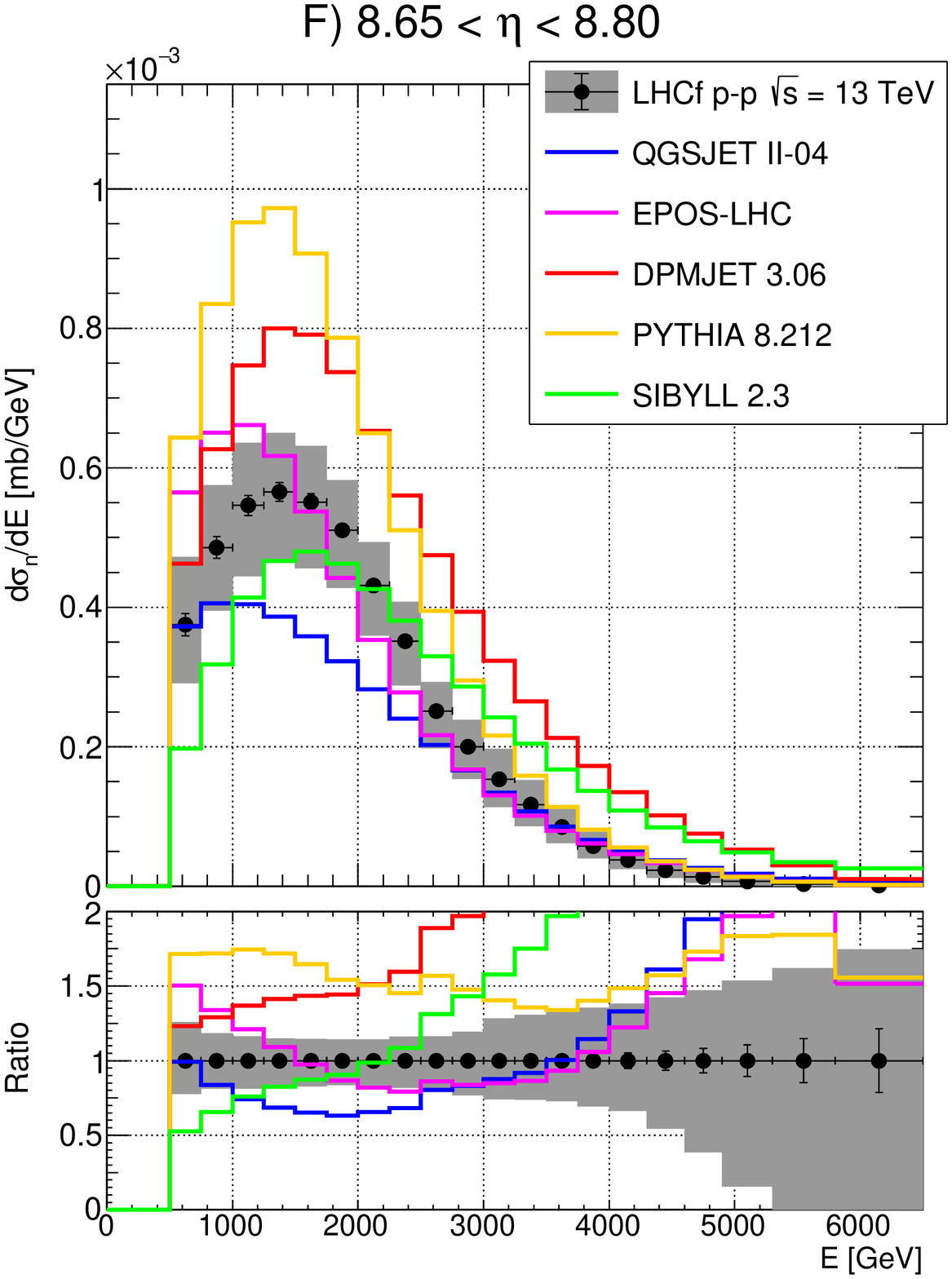}%
 \caption{Inclusive differential neutron production cross section for p-p collisions at $\sqrt{s} = 13$~TeV, measured using the LHCf Arm2 detector. Black markers represent the experimental data with statistical errors, whereas gray bands represent the quadratic sum of statistical and systematic uncertainties. Colored histograms refer to model predictions at the generator level. For each region, the top plot shows the energy distributions expressed as $\mathrm{d}\sigma_{\mathrm{n}}/\mathrm{d}E$ and the bottom plot the ratios of these distributions to the experimental results.}
\label{fig:unfolded}
\end{figure*}

\enlargethispage{-1\baselineskip}
In this section, the measurement of neutron production is presented and the comparison with model predictions is discussed. Distributions are expressed in terms of the inclusive differential production cross section $\mathrm{d}\sigma_{\mathrm{n}}/\mathrm{d}E$ for six pseudorapidity intervals. In each interval, the distribution is corrected to take into account the limited coverage of the detector in terms of the azimuthal angle. The data are then compared with the generator predictions, using for each model its own inelastic cross section.  

The $\mathrm{d}\sigma_{\mathrm{n}}/\mathrm{d}E$ unfolded distributions are shown in figure \ref{fig:unfolded}. The measurements are limited to energies above $\mathrm{500~GeV}$ because, as described in section \ref{sec:reconstruction}, the hardware trigger was optimized to have a good detection efficiency above this value. The total uncertainty is given by the quadratic sum of all statistical and systematic sources: the systematic uncertainties are the dominant contribution everywhere, whereas the statistical uncertainties are generally smaller than $5\%$. The numerical values of the inclusive differential production cross section are also summarized in appendix \ref{app:cross_section}. 

At a first glance, it is interesting to note that all experimental distributions exhibit a peak structure, whose position progressively moves to lower energy as pseudorapidity decreases, from about $\mathrm{5~TeV}$ for $\eta > 10.75$ to about $\mathrm{1.5~TeV}$ for $8.65 < \eta < 8.80$. Furthermore, it is found that all these distributions, except the one relative to the most forward region, can be fitted by a very simple function like an asymmetric gaussian -- a fact that could give a hint on the underlying physical process responsible for neutron production.

Considering the data-model comparison, the largest deviation between model distributions and experimental data is at $\eta > 10.75$, as already noted in \cite{ref:LHCf_Neutron}. In this region, LHCf measurements indicate a peak structure at around 5~TeV which is not predicted by any generator and, in addition, all models underestimate the total production cross section by at least $20\%$ (QGSJET II-04). No model completely reproduces the experimental measurements even in the other five regions, although deviations are smaller than at $\eta > 10.75$. Depending on the pseudorapidity region, the generators showing the best overall agreement with data are either SIBYLL 2.3 or EPOS-LHC. At $10.06 < \eta < 10.75$, EPOS-LHC reproduces well the peak position but it underestimates neutron production, whereas, at $9.65 < \eta < 10.06$, SIBYLL 2.3 reproduces well the peak position but it overestimates neutron production. In all the remaining three regions, it is interesting to note that SIBYLL 2.3 is compatible with the experimental measurements in the energy range between 1.5 TeV and 2.5~TeV, where the production peak is located. However, SIBYLL 2.3 is the model showing the best overall agreement with data only at $8.99 < \eta < 9.21$, whereas EPOS-LHC leads to better results at $8.80 < \eta < 8.99$ and $8.65 < \eta < 8.80$.

\section{Discussion}
\label{sec:discussion}

\enlargethispage{+1\baselineskip}		
The results presented in section \ref{sec:results} are interesting by themselves and useful for model tuning. However, in order to test the validity of generators in the context of cosmic ray measurements, it is better to express these results in terms of quantities that are directly connected to air shower physics. In this section, the distributions shown in figure \ref{fig:unfolded} are therefore used to extract three important parameters: energy flow, cross section and average inelasticity of forward neutrons.

The differential energy flow $\mathrm{d}E_{\mathrm{n}}/\mathrm{d}\eta$ and the differential cross section $\mathrm{d}\sigma_{\mathrm{n}}/\mathrm{d}\eta$ are expressed as a function of pseudorapidity in the following manner. For each region, the corresponding mean pseudorapidity, $\eta$, and pseudorapidity interval, $\Delta \eta$, are given by the average value and the distance of the two extremes, respectively\footnote{
In case of the most forward region, the upper limit of $\infty$ is limited to 13 for computational reasons. This number was chosen in such a way that more than 95\% of the events in this region are below this value.
}. In a similar way, for each energy bin $i$ of the $\mathrm{d}\sigma_{\mathrm{n}}/\mathrm{d}E$ distribution, the mean energy, $E_{i}$, and the energy interval, $\mathrm{d}E_{i}$, are given by the average value and the distance of the two extremes, respectively. Thus, $\mathrm{d}E_{\mathrm{n}}/\mathrm{d}\eta$ and $\mathrm{d}\sigma_{\mathrm{n}}/\mathrm{d}\eta$ are given by
\begin{equation*}
\dfrac{\mathrm{d}E_{\mathrm{n}}}{\mathrm{d}\eta} = \dfrac{1}{\Delta \eta} \dfrac{1}{\sigma_{inel}} 
\sum_{i} \dfrac{\mathrm{d}\sigma_{\mathrm{n}}}{\mathrm{d}E}\bigg\vert_{i} E_{i} \mathrm{d}E_{i}
 \quad\quad\mathrm{and}\quad\quad
\dfrac{\mathrm{d}\sigma_{\mathrm{n}}}{\mathrm{d}\eta} = \dfrac{1}{\Delta \eta}
\sum_{i} \dfrac{\mathrm{d}\sigma_{\mathrm{n}}}{\mathrm{d}E}\bigg\vert_{i} \mathrm{d}E_{i}
\end{equation*}
where the TOTEM measurement of $\mathrm{\sigma_{inel} = (79.5 \pm 1.8) ~ mb}$ was used for the total inelastic cross section (and the corresponding uncertainty) of p-p collisions at $\mathrm{\sqrt{s} = 13~TeV}$  \cite{ref:TOTEM_CrossSection}. Note that these quantities must be corrected to take into account the contribution of neutrons below $\mathrm{500~GeV}$, which are not included in the $\mathrm{d}\sigma_{\mathrm{n}}/\mathrm{d}E$ distributions. Correction factors were estimated from five simulation samples generated using all the models discussed in section \ref{sec:mc}, taking the average as best estimation and the maximum deviation as uncertainty. For $\mathrm{d}E_{\mathrm{n}}/\mathrm{d}\eta$, the corrections are at most 1.5\% and absolute uncertainties below 1\%, whereas, for $\mathrm{d}\sigma_{\mathrm{n}}/\mathrm{d}\eta$, the corrections are at most 8\% and absolute uncertainties below 5\%. Another important aspect is how the several sources of uncertainty acting on $\mathrm{d}\sigma_{\mathrm{n}}/\mathrm{d}E$ contribute to the uncertainty of these two derived quantities. First, all contributions are assumed to be independent, so that they can be added in quadrature. Then, the uncertainties are divided in two groups: bin-by-bin independent (only statistical) and bin-by-bin fully-correlated (all systematic). For each bin-by-bin fully-correlated source, its contribution to the final uncertainty is given by the maximum deviation between the quantity derived from the nominal distribution and the one derived from the distributions corresponding to all possible shifts induced by that uncertainty. 

\begin{figure*}[tbp]
 \centering
 \includegraphics[width=0.45\textwidth]{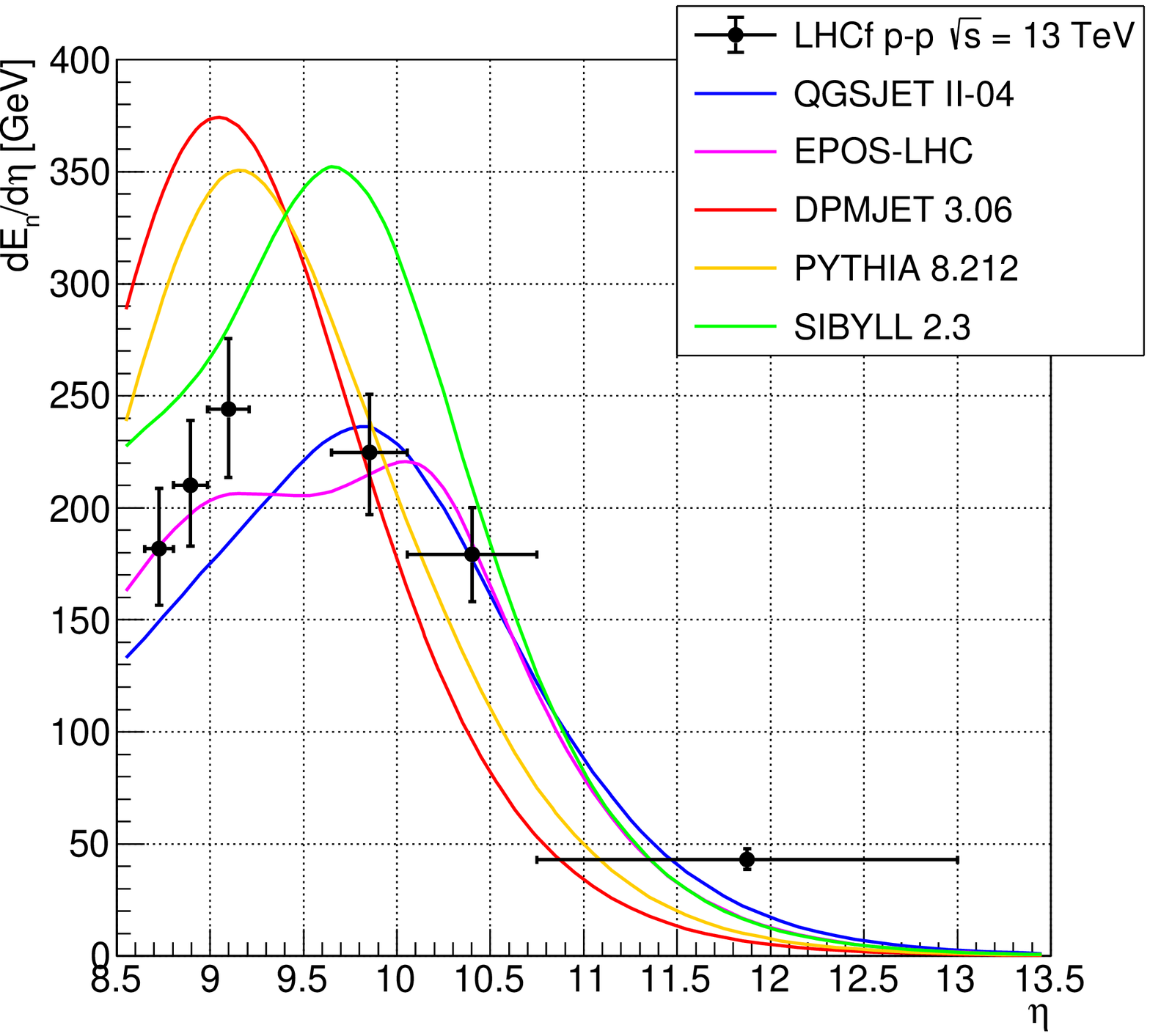}%
 \includegraphics[width=0.45\textwidth]{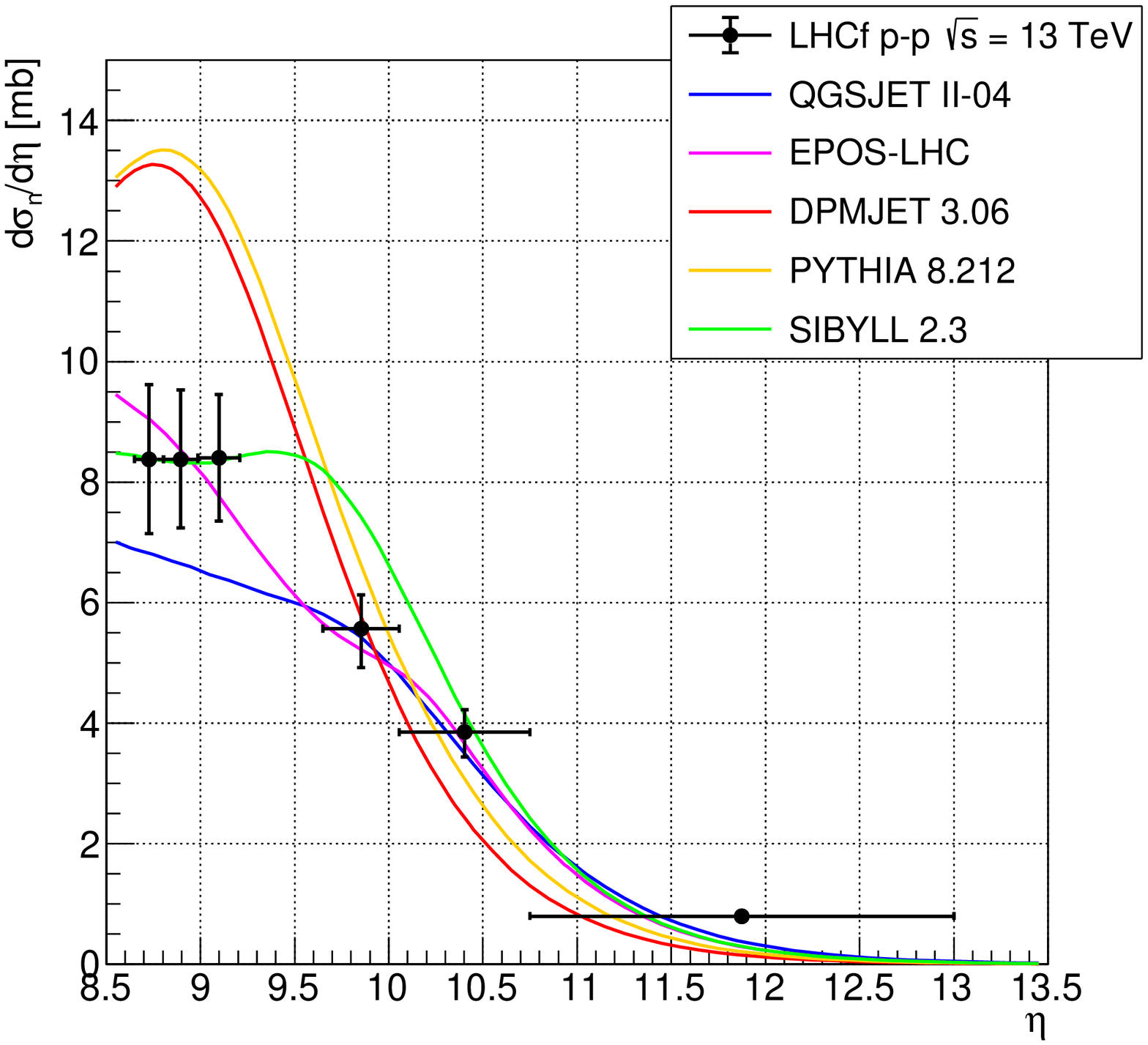}%
 \caption{Differential energy flow $\mathrm{d}E_{\mathrm{n}}/\mathrm{d}\eta$ (left) and differential cross section $\mathrm{d}\sigma_{\mathrm{n}}/\mathrm{d}\eta$ (right) of neutrons produced in p-p collisions at $\sqrt{s} = 13$~TeV, measured using the LHCf Arm2 detector. Black markers represent the experimental data with statistical and systematic uncertainties, whereas colored lines refer to model predictions at the generator level.}
\label{fig:energy_flow_and_cross_section}
\end{figure*}

Figure \ref{fig:energy_flow_and_cross_section} shows the differential energy flow and the differential production cross section measured using the LHCf Arm2 detector. The numerical values of $\mathrm{d}\sigma_{\mathrm{n}}/\mathrm{d}\eta$ and $\mathrm{d}E_{\mathrm{n}}/\mathrm{d}\eta$ are also summarized in table \ref{tab:energy_flow_and_cross_section}.  As expected, in both cases all models underestimate these quantities at $\eta > 10.75$, whereas for the other regions the disagreement between generators and data is less serious. In particular, EPOS-LHC is the model showing the best overall agreement with the experimental measurements, QGSJET II-04 is consistent with data at $9.65 < \eta < 10.75$, but underestimates these quantities at $8.65 < \eta < 9.21$, whereas SIBYLL 2.3 reasonably reproduces $\mathrm{d}\sigma_{\mathrm{n}}/\mathrm{d}\eta$, but strongly overestimates $\mathrm{d}E_{\mathrm{n}}/\mathrm{d}\eta$. Furthermore, even if there are no data points for $\eta$ values between $9.21$ and $9.65$, experimental measurements seem to indicate that the peak in $\mathrm{d}E_{\mathrm{n}}/\mathrm{d}\eta$ must be located here, if this distribution has such kind of structure as predicted by all models except EPOS-LHC.

\begin{table}[tbp]
  \begin{center}
    \begin{tabular}{|P{3cm}|P{2.6cm}|P{2.6cm}|P{2.3cm}|P{2.3cm}|}
      \hline
	  & $\mathrm{d}E_{\mathrm{n}}/\mathrm{d}\eta~\mathrm{[GeV]}$ ($E > 500~\mathrm{GeV}$) 
	  & $\mathrm{d}\sigma_{\mathrm{n}}/\mathrm{d}\eta~\mathrm{[mb]}$ ($E > 500~\mathrm{GeV}$) 
	  & $\mathrm{d}E_{\mathrm{n}}/\mathrm{d}\eta~\mathrm{[GeV]}$ ($E > 0~\mathrm{GeV}$) 
	  & $\mathrm{d}\sigma_{\mathrm{n}}/\mathrm{d}\eta~\mathrm{[mb]}$ ($E > 0~\mathrm{GeV}$) \\
	  \hline
	  $ 8.65 < \eta <  8.80$ & $179.6_{-24.9}^{+26.6}$ & $7.77_{-1.08}^{+1.10}$  
	  					     & $181.8_{-25.2}^{+27.0}$ & $8.38_{-1.23}^{+1.24}$ \\
	  $ 8.80 < \eta <  8.99$ & $208.4_{-26.8}^{+28.7}$ & $7.92_{-1.03}^{+1.05}$
	  					     & $210.1_{-27.1}^{+29.0}$ & $8.38_{-1.13}^{+1.15}$ \\
	  $ 8.99 < \eta <  9.21$ & $242.7_{-30.2}^{+31.5}$ & $8.07_{-0.99}^{+0.99}$
					  	     & $244.0_{-30.4}^{+31.7}$ & $8.40_{-1.05}^{+1.05}$ \\
	  $~9.65 < \eta < 10.06$ & $224.4_{-27.7}^{+26.0}$ & $5.49_{-0.64}^{+0.55}$
					         & $224.7_{-27.7}^{+26.1}$ & $5.57_{-0.65}^{+0.56}$ \\
	  $10.06 < \eta < 10.75$ & $179.0_{-21.0}^{+21.0}$ & $3.82_{-0.41}^{+0.37}$
						     & $179.2_{-21.0}^{+21.0}$ & $3.85_{-0.41}^{+0.38}$ \\
	  $\eta > 10.75$ 		 & $43.0_{-4.3}^{+4.8}$ & $0.79_{-0.07}^{+0.07}$ 
	  						 & $43.0_{-4.3}^{+4.8}$ & $0.80_{-0.07}^{+0.08}$ \\
      \hline
    \end{tabular}
    \caption{Differential energy flow $\mathrm{d}E_{\mathrm{n}}/\mathrm{d}\eta$ and differential cross section $\mathrm{d}\sigma_{\mathrm{n}}/\mathrm{d}\eta$ of neutrons produced in p-p collisions at $\sqrt{s} = 13$~TeV, measured using the LHCf Arm2 detector. Upper and lower uncertainties are also reported. The values are relative to the experimental measurements \textit{with} ($E > 0~\mathrm{GeV}$) and \textit{without} ($E > 500~\mathrm{GeV}$) the simulation-driven correction factors for the limited detection efficiency below $\mathrm{500~GeV}$. The last two columns correspond to the numbers used for the experimental points shown in figure \ref{fig:energy_flow_and_cross_section}. }
    \label{tab:energy_flow_and_cross_section}
  \end{center}
\end{table}

The average inelasticity $\langle 1-k \rangle$ can be extracted from the elasticity distribution, where the elasticity is defined as $k = 2E / \sqrt{s}$, and $E$ is the energy of the leading particle. The \textit{leading particle} is the particle (not necessarily a baryon) that, on one of the two sides of the collision (let us say $\eta>0$), carries the largest fraction of the primary proton momentum. Note that, being sensitive to neutrons only, the LHCf measurements are restricted to the specific case where the leading particle is a neutron. To distinguish the quantities that are measured here from their general definition, in the following they are indicated as $k_{\mathrm{n}}$ and $\langle 1-k_{\mathrm{n}} \rangle$, instead of $k$ and $\langle 1-k \rangle$. The role of neutrons as forward leading particles was investigated with the use of generators. Despite the large variations among the different models, two important features were clearly highlighted. The first one is that, depending on the energy bin, the fraction of events where the leading particle is a neutron ranges from about $5$ to $35\%$. The second one is that, for energies above half the beam energy, almost $100\%$ of the neutrons produced from the collisions are leading particles. In order to obtain the elasticity distribution, the $\mathrm{d}\sigma_{\mathrm{n}}/\mathrm{d}E$ contributions of all the six regions are summed in a single histogram. Then, the x axis is rescaled to the beam energy and the y axis is multiplied for the bin width, so that the distribution represents the total production cross section $\sigma_{\mathrm{n}}$ as a function of elasticity $k_{\mathrm{n}}$. At this point, a correction must be applied to take into account two different effects: the first one is due to the fact that the detector has a limited pseudorapidity coverage; the second one is due to the fact that not all neutrons are leading particles. These two effects are considered together in a single correction factor that was obtained from five simulation samples generated using all the models discussed in section \ref{sec:mc}, taking the average as best estimate and the maximum deviation as uncertainty. Corrections range between 1\% and 70\%, whereas absolute uncertainties go from 5\% to 70\%. The several sources of uncertainties acting on the $\mathrm{d}\sigma_{\mathrm{n}}/\mathrm{d}E$ distributions contribute to the uncertainty on the elasticity distribution in a similar way to what was previously described, \textit{i.e.} assuming that all contributions are independent and dividing them in bin-by-bin independent (only statistical) and bin-by-bin fully-correlated (all systematic) sources. Note that, differently from the previous case, the term \textit{bin} does not refer to the energy bin, but to the pseudorapidity bin, because the summation index is on pseudorapidity and not on energy. The elasticity distribution cannot directly be used to extract the error on the average inelasticity, because systematic uncertainties are correlated both on energy and on pseudorapidity. The entire procedure must therefore be repeated to extract the uncertainty, but an average value is computed instead of building a histogram, so that both sources of correlation are correctly considered in the estimation of the uncertainty. This value is then corrected to take into account the contribution of neutrons below $\mathrm{500~GeV}$, which are not included in the $\mathrm{d}\sigma_{\mathrm{n}}/\mathrm{d}E$ distributions. The correction factor, estimated in a similar way to what was previously discussed, amounts to a value of $(0.4 \pm 0.4) \%$.

\begin{figure*}[tbp]
 \centering
 \includegraphics[width=0.45\textwidth]{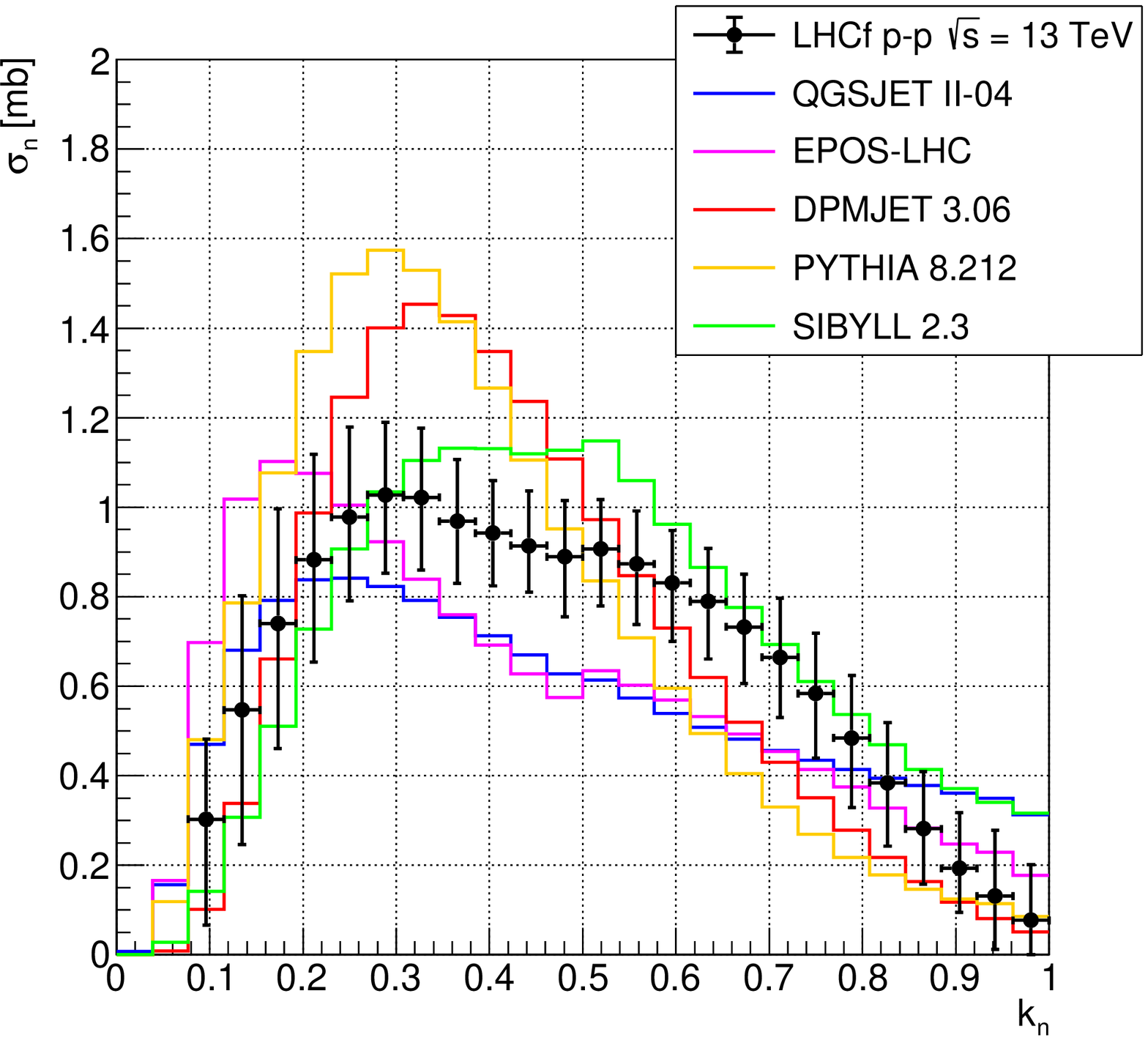}%
 \includegraphics[width=0.45\textwidth]{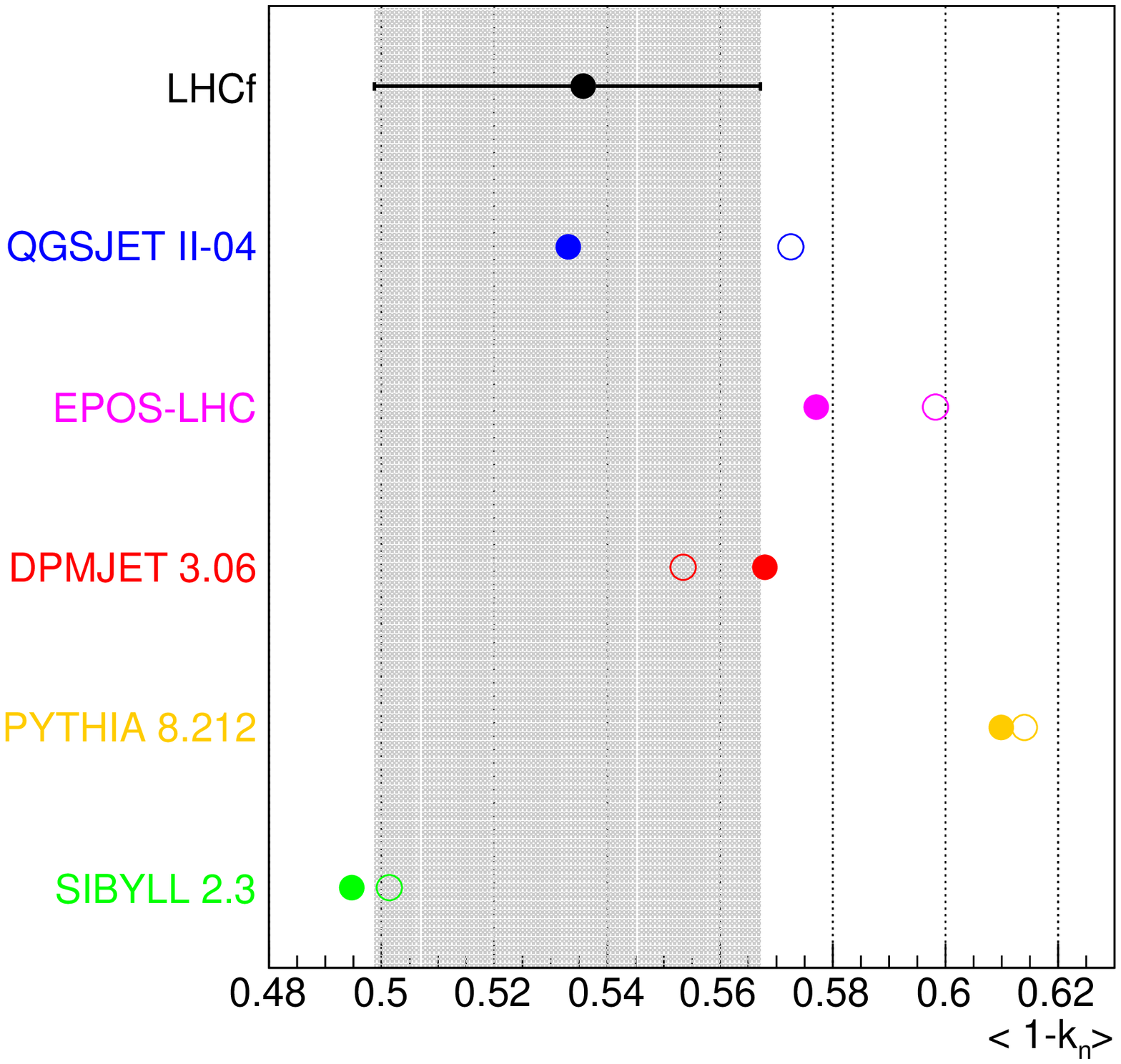}%
 \caption{Inclusive production cross section as a function of elasticity $k_{\mathrm{n}}$ (left) and average inelasticity $\langle 1-k_{\mathrm{n}} \rangle$ extracted from that distribution (right), relative to p-p collisions at $\sqrt{s} = 13$~TeV. These quantities, measured using the LHCf Arm2 detector, are only relative to the events where the leading particle is a neutron. Black markers represent the experimental data with the quadratic sum of statistical and systematic uncertainties. Solid lines (left) and full circles (right) refer to model predictions at the generator level, obtained using only the events where the leading particle is a neutron. In order to compare this approach to the general case, $\langle 1-k \rangle$, the average inelasticity obtained using all the events independently of the nature of the leading particle, is also reported as open circles in the right figure.
 }
\label{fig:elasticity}
\end{figure*}

Figure \ref{fig:elasticity} shows the inclusive production cross section as a function of elasticity and the average inelasticity extracted from that distribution, measured using the LHCf Arm2 detector. In the left plot, the contribution to the error bars of $\sigma_{\mathrm{n}}$ is dominated by the uncertainty on $\mathrm{d}\sigma_{\mathrm{n}}/\mathrm{d}E$ for large values of $k_{\mathrm{n}}$ and by the uncertainty on elasticity correction factors for small values of $k_{\mathrm{n}}$. As can be seen, no generator shows a satisfactory agreement with the experimental measurements for all the elasticity range. In particular, for $0.3 < k_{\mathrm{n}} < 0.6$, the data lies between SIBYLL 2.3 and EPOS-LHC, QGSJET II-04 predictions. Outside this region, the model most consistent with the data is SIBYLL 2.3 for $k_{\mathrm{n}} < 0.3$, $0.6 < k_{\mathrm{n}} < 0.8$ and EPOS-LHC for $k_{\mathrm{n}} > 0.8$. The average inelasticity extracted from the elasticity distribution with the method just discussed is $\langle 1-k_{\mathrm{n}} \rangle = (0.536^{+0.031}_{-0.037})$. In the right plot, this value is compared to model predictions for two different quantities: $\langle 1-k_{\mathrm{n}} \rangle$ (experimental case) and $\langle 1-k \rangle$ (general case). As can be seen, all generators predict similar numbers for these two quantities, indicating that, even if limited only to the events where the leading particle is a neutron, this measurement provides a relevant information also for the general definition of inelasticity. With a $\langle 1-k_{\mathrm{n}} \rangle$ value of 0.533, QGSJET II-04 is the only model consistent with the experimental measurement, even if all the remaining generators except PYTHIA 8.212 lie just outside the error bars. As a final remark, it is important to stress that, despite the relative agreement found on the average inelasticity, the elasticity distribution predicted by all models differs from the one observed in the data.

\section{Summary}

The LHCf experiment measured the inclusive differential production cross section of forward neutrons (+ antineutrons) produced in $\sqrt{s} = 13$~TeV proton-proton collisions. The analysis covers the following six pseudorapidity regions: $\eta > 10.75$, $10.06 < \eta < 10.75$, $9.65 < \eta < 10.06$, $8.99 < \eta < 9.21$, $8.80 < \eta < 8.99$ and $8.65 < \eta < 8.80$. The LHCf measurements were compared to the predictions of several hadronic interaction models: QGSJET II-04, EPOS-LHC, SIBYLL 2.3, DPMJET 3.06 and PYTHIA 8.212. All generators dramatically differ from the experimental results for $\eta > 10.75$, whereas at lower pseudorapidities the data-model agreement is better, but still not satisfactory. In this last case, depending on the pseudorapidity region, the generator showing the best overall agreement with data is either SIBYLL 2.3 or EPOS-LHC. From the inclusive differential production cross section, three quantities directly connected to air shower development were derived: neutron energy flow, cross section and average inelasticity. The first and the second quantities are well reproduced by EPOS-LHC, except for the most forward region, whereas QGSJET II-04 gives a consistent value with the data for the third quantity, with most generators lying just outside the experimental uncertainties. However, despite the relative agreement observed on the average inelasticity of leading neutrons, no generator satisfactorily reproduces the corresponding elasticity distribution. 

The LHCf results are the first measurements of forward neutron production at such a high energy and, given the significant deviation between generators and data, they suggest that all models would benefit from a tuning that takes these results into account. A further development of this analysis is possible thanks to the fact that the LHCf and ATLAS experiments had common data acquisition during the operations where the data considered here was recorded. By exploiting the ATLAS information in the central region, it will be possible to investigate the different mechanisms responsible for neutron production in the forward region \cite{ref:Diffractive_Zhou} and the amount of correlation between central and forward hadron production \cite{ref:CentralForward}.

\acknowledgments
We thank the CERN staff and the ATLAS Collaboration for their essential contributions to the successful operation of LHCf. We are grateful to S. Ostapchenko for useful comments about QGSJET II-04 generator and to the developers of CRMC interface tool for its implementation. This work was supported by several institutions in Japan and in Italy: in Japan, by the Japanese Society for the Promotion of Science (JSPS) KAKENHI (Grant Numbers JP26247037, JP23340076) and the joint research program of the Institute for Cosmic Ray Research (ICRR), University of Tokyo; in Italy, by Istituto Nazionale di Fisica Nucleare (INFN) and by the University of Catania (Grant Numbers UNICT 2020-22 Linea 2). This work took advantage of computer resources supplied by ICRR (University of Tokyo), CERN and CNAF (INFN).

\clearpage

\appendix
\section{Table of inclusive differential neutron production cross section}
\label{app:cross_section}

\begin{table} [htbp]
  \scriptsize
  \begin{center}
    \begin{tabular}{|P{1.5cm}|P{2.65cm}|P{1.5cm}|P{2.65cm}|P{1.5cm}|P{2.65cm}|}
      \hline
      \multicolumn{2}{|c}{} & \multicolumn{2}{|c|}{} & \multicolumn{2}{c|}{} \\
      \multicolumn{2}{|c}{\boldmath$\eta > 10.75$} & \multicolumn{2}{|c|}{\boldmath$10.06 < \eta < 10.75$} & \multicolumn{2}{c|}{\boldmath$9.65 < \eta < 10.06$} \\
      \multicolumn{2}{|c}{} & \multicolumn{2}{|c|}{} & \multicolumn{2}{c|}{} \\
      \hline
      &  &  &  &  &\\
      \textbf{Energy [GeV]} & \boldmath$\mathrm{d}\sigma_{\mathrm{n}}/\mathrm{d}E$ \textbf{[mb/GeV]} & \textbf{Energy [GeV]} & \boldmath$\mathrm{d}\sigma_{\mathrm{n}}/\mathrm{d}E$ \textbf{[mb/GeV]}
      & \textbf{Energy [GeV]} & \boldmath$\mathrm{d}\sigma_{\mathrm{n}}/\mathrm{d}E$ \textbf{[mb/GeV]} \\
      &  &  &  &  &\\
      \hline
      &  &  &  &  &\\
      500--750 & $(2.09_{-1.13}^{+1.03}) \times 10^{-5}$  &       500--750 & $(8.66_{-3.59}^{+3.45}) \times 10^{-5}$  &       500--750 & $(1.35_{-0.52}^{+0.61}) \times 10^{-4}$ \\
      750--1000 & $(5.12_{-1.77}^{+1.67}) \times 10^{-5}$  &       750--1000 & $(1.35_{-0.42}^{+0.39}) \times 10^{-4}$  &       750--1000 & $(2.16_{-0.65}^{+0.57}) \times 10^{-4}$ \\
      1000--1250 & $(7.84_{-2.21}^{+1.91}) \times 10^{-5}$  &       1000--1250 & $(1.86_{-0.47}^{+0.41}) \times 10^{-4}$  &       1000--1250 & $(2.49_{-0.65}^{+0.48}) \times 10^{-4}$ \\
      1250--1500 & $(9.75_{-2.13}^{+2.03}) \times 10^{-5}$  &       1250--1500 & $(2.36_{-0.47}^{+0.42}) \times 10^{-4}$  &       1250--1500 & $(2.82_{-0.63}^{+0.50}) \times 10^{-4}$ \\
      1500--1750 & $(1.09_{-0.21}^{+0.23}) \times 10^{-4}$  &       1500--1750 & $(2.77_{-0.50}^{+0.45}) \times 10^{-4}$  &       1500--1750 & $(3.27_{-0.67}^{+0.56}) \times 10^{-4}$ \\
      1750--2000 & $(1.15_{-0.20}^{+0.25}) \times 10^{-4}$  &       1750--2000 & $(3.06_{-0.51}^{+0.48}) \times 10^{-4}$  &       1750--2000 & $(3.78_{-0.74}^{+0.60}) \times 10^{-4}$ \\
      2000--2250 & $(1.24_{-0.22}^{+0.31}) \times 10^{-4}$  &       2000--2250 & $(3.41_{-0.54}^{+0.52}) \times 10^{-4}$  &       2000--2250 & $(4.40_{-0.84}^{+0.65}) \times 10^{-4}$ \\
      2250--2500 & $(1.39_{-0.24}^{+0.25}) \times 10^{-4}$  &       2250--2500 & $(3.82_{-0.62}^{+0.60}) \times 10^{-4}$  &       2250--2500 & $(5.03_{-0.96}^{+0.72}) \times 10^{-4}$ \\
      2500--2750 & $(1.60_{-0.29}^{+0.25}) \times 10^{-4}$  &       2500--2750 & $(4.38_{-0.76}^{+0.72}) \times 10^{-4}$  &       2500--2750 & $(5.66_{-1.04}^{+0.79}) \times 10^{-4}$ \\
      2750--3000 & $(1.93_{-0.39}^{+0.34}) \times 10^{-4}$  &       2750--3000 & $(5.06_{-0.89}^{+0.94}) \times 10^{-4}$  &       2750--3000 & $(6.19_{-1.07}^{+0.88}) \times 10^{-4}$ \\
      3000--3250 & $(2.26_{-0.51}^{+0.46}) \times 10^{-4}$  &       3000--3250 & $(5.93_{-1.04}^{+1.14}) \times 10^{-4}$  &       3000--3250 & $(6.60_{-1.11}^{+0.94}) \times 10^{-4}$ \\
      3250--3500 & $(2.87_{-0.76}^{+0.68}) \times 10^{-4}$  &       3250--3500 & $(6.87_{-1.21}^{+1.32}) \times 10^{-4}$  &       3250--3500 & $(6.79_{-1.09}^{+0.90}) \times 10^{-4}$ \\
      3500--3750 & $(3.51_{-0.86}^{+0.90}) \times 10^{-4}$  &       3500--3750 & $(7.58_{-1.33}^{+1.46}) \times 10^{-4}$  &       3500--3750 & $(6.75_{-1.19}^{+0.87}) \times 10^{-4}$ \\
      3750--4000 & $(4.35_{-1.02}^{+1.18}) \times 10^{-4}$  &       3750--4000 & $(8.19_{-1.46}^{+1.50}) \times 10^{-4}$  &       3750--4000 & $(6.26_{-1.23}^{+0.82}) \times 10^{-4}$ \\
      4000--4250 & $(5.25_{-1.23}^{+1.38}) \times 10^{-4}$  &       4000--4250 & $(8.44_{-1.47}^{+1.41}) \times 10^{-4}$  &       4000--4250 & $(5.63_{-1.31}^{+0.84}) \times 10^{-4}$ \\
      4250--4500 & $(5.99_{-1.39}^{+1.39}) \times 10^{-4}$  &       4250--4500 & $(8.19_{-1.30}^{+1.30}) \times 10^{-4}$  &       4250--4500 & $(4.88_{-1.34}^{+0.96}) \times 10^{-4}$ \\
      4500--4750 & $(6.45_{-1.40}^{+1.20}) \times 10^{-4}$  &       4500--4750 & $(7.65_{-1.27}^{+1.33}) \times 10^{-4}$  &       4500--4750 & $(4.07_{-1.33}^{+1.07}) \times 10^{-4}$ \\
      4750--5000 & $(6.59_{-1.19}^{+1.09}) \times 10^{-4}$  &       4750--5000 & $(6.75_{-1.43}^{+1.41}) \times 10^{-4}$  &       4750--5000 & $(3.28_{-1.25}^{+1.13}) \times 10^{-4}$ \\
      5000--5250 & $(6.36_{-1.08}^{+1.21}) \times 10^{-4}$  &       5000--5250 & $(5.44_{-1.50}^{+1.56}) \times 10^{-4}$  &       5000--5250 & $(2.52_{-1.13}^{+1.09}) \times 10^{-4}$ \\
      5250--5500 & $(5.54_{-1.36}^{+1.41}) \times 10^{-4}$  &       5250--5500 & $(4.34_{-1.51}^{+1.61}) \times 10^{-4}$  &       5250--5500 & $(1.85_{-0.95}^{+0.99}) \times 10^{-4}$ \\
      5500--5750 & $(4.44_{-1.71}^{+1.55}) \times 10^{-4}$  &       5500--5750 & $(3.12_{-1.36}^{+1.61}) \times 10^{-4}$  &       5500--5750 & $(1.27_{-0.75}^{+0.75}) \times 10^{-4}$ \\
      5750--6000 & $(3.12_{-1.56}^{+1.86}) \times 10^{-4}$  &       5750--6000 & $(2.17_{-1.13}^{+1.44}) \times 10^{-4}$  &       5750--6000 & $(9.37_{-6.89}^{+6.20}) \times 10^{-5}$ \\
      6000--6250 & $(2.37_{-1.86}^{+2.12}) \times 10^{-4}$  &       6000--6250 & $(1.37_{-1.12}^{+1.27}) \times 10^{-4}$  &       6000--6250 & $(6.27_{-6.27}^{+5.05}) \times 10^{-5}$ \\
      6250--6500 & $(1.40_{-1.40}^{+1.92}) \times 10^{-4}$  &       6250--6500 & $(8.01_{-8.01}^{+10.05}) \times 10^{-5}$  &       6250--6500 & $(4.03_{-4.03}^{+4.01}) \times 10^{-5}$ \\
      &  &  &  &  &\\
      \hline
    \end{tabular}
    \caption{Inclusive differential neutron production cross section for p-p collisions at $\sqrt{s} = 13$~TeV, relative to three of the six regions used in the analysis. These regions are located on the  small tower of the LHCf Arm2 detector. Upper and lower uncertainties, expressed as the quadratic sum of statistical and systematics contributions, are also reported.}
    \label{tab:results_small}
  \end{center}
\end{table}

\begin{table} [htbp]
  \scriptsize
  \begin{center}
    \begin{tabular}{|P{1.5cm}|P{2.65cm}|P{1.5cm}|P{2.65cm}|P{1.5cm}|P{2.65cm}|}
      \hline
      \multicolumn{2}{|c}{} & \multicolumn{2}{|c|}{} & \multicolumn{2}{c|}{} \\
      \multicolumn{2}{|c}{\boldmath$8.99 < \eta < 9.21$} & \multicolumn{2}{|c|}{\boldmath$8.80 < \eta < 8.99$} & \multicolumn{2}{c|}{\boldmath$8.65 < \eta < 8.80$} \\
      \multicolumn{2}{|c}{} & \multicolumn{2}{|c|}{} & \multicolumn{2}{c|}{} \\
      \hline
      &  &  &  &  &\\
      \textbf{Energy [GeV]} & \boldmath$\mathrm{d}\sigma_{\mathrm{n}}/\mathrm{d}E$ \textbf{[mb/GeV]} & \textbf{Energy [GeV]} & \boldmath$\mathrm{d}\sigma_{\mathrm{n}}/\mathrm{d}E$ \textbf{[mb/GeV]}
      & \textbf{Energy [GeV]} & \boldmath$\mathrm{d}\sigma_{\mathrm{n}}/\mathrm{d}E$ \textbf{[mb/GeV]} \\
      &  &  &  &  &\\
      \hline
      &  &  &  &  &\\
      500--750 & $(2.70_{-0.69}^{+0.96}) \times 10^{-4}$  &       500--750 & $(3.16_{-0.79}^{+0.96}) \times 10^{-4}$  &       500--750 & $(3.75_{-0.84}^{+0.97}) \times 10^{-4}$ \\
      750--1000 & $(3.81_{-0.82}^{+0.92}) \times 10^{-4}$  &       750--1000 & $(4.45_{-0.95}^{+0.96}) \times 10^{-4}$  &       750--1000 & $(4.86_{-0.90}^{+0.89}) \times 10^{-4}$ \\
      1000--1250 & $(4.73_{-0.97}^{+0.84}) \times 10^{-4}$  &       1000--1250 & $(5.49_{-1.15}^{+1.03}) \times 10^{-4}$  &       1000--1250 & $(5.46_{-1.01}^{+0.90}) \times 10^{-4}$ \\
      1250--1500 & $(5.40_{-0.94}^{+0.89}) \times 10^{-4}$  &       1250--1500 & $(5.91_{-1.21}^{+1.14}) \times 10^{-4}$  &       1250--1500 & $(5.66_{-1.02}^{+0.84}) \times 10^{-4}$ \\
      1500--1750 & $(5.97_{-1.00}^{+0.93}) \times 10^{-4}$  &       1500--1750 & $(5.91_{-1.11}^{+1.16}) \times 10^{-4}$  &       1500--1750 & $(5.51_{-0.94}^{+0.80}) \times 10^{-4}$ \\
      1750--2000 & $(6.39_{-1.05}^{+0.97}) \times 10^{-4}$  &       1750--2000 & $(5.71_{-1.03}^{+1.13}) \times 10^{-4}$  &       1750--2000 & $(5.10_{-0.82}^{+0.72}) \times 10^{-4}$ \\
      2000--2250 & $(6.35_{-1.03}^{+0.96}) \times 10^{-4}$  &       2000--2250 & $(5.15_{-0.90}^{+1.07}) \times 10^{-4}$  &       2000--2250 & $(4.31_{-0.72}^{+0.62}) \times 10^{-4}$ \\
      2250--2500 & $(5.60_{-0.77}^{+0.83}) \times 10^{-4}$  &       2250--2500 & $(4.35_{-0.74}^{+0.94}) \times 10^{-4}$  &       2250--2500 & $(3.51_{-0.63}^{+0.56}) \times 10^{-4}$ \\
      2500--2750 & $(5.70_{-0.76}^{+0.84}) \times 10^{-4}$  &       2500--2750 & $(3.68_{-0.67}^{+0.79}) \times 10^{-4}$  &       2500--2750 & $(2.51_{-0.53}^{+0.41}) \times 10^{-4}$ \\
      2750--3000 & $(5.30_{-0.69}^{+0.77}) \times 10^{-4}$  &       2750--3000 & $(3.25_{-0.64}^{+0.75}) \times 10^{-4}$  &       2750--3000 & $(2.00_{-0.46}^{+0.39}) \times 10^{-4}$ \\
      3000--3250 & $(4.74_{-0.67}^{+0.71}) \times 10^{-4}$  &       3000--3250 & $(2.72_{-0.58}^{+0.67}) \times 10^{-4}$  &       3000--3250 & $(1.54_{-0.40}^{+0.43}) \times 10^{-4}$ \\
      3250--3500 & $(3.98_{-0.61}^{+0.65}) \times 10^{-4}$  &       3250--3500 & $(2.22_{-0.49}^{+0.61}) \times 10^{-4}$  &       3250--3500 & $(1.17_{-0.30}^{+0.36}) \times 10^{-4}$ \\
      3500--3750 & $(3.20_{-0.57}^{+0.62}) \times 10^{-4}$  &       3500--3750 & $(1.71_{-0.41}^{+0.52}) \times 10^{-4}$  &       3500--3750 & $(8.50_{-2.30}^{+2.90}) \times 10^{-5}$ \\
      3750--4000 & $(2.38_{-0.49}^{+0.58}) \times 10^{-4}$  &       3750--4000 & $(1.26_{-0.33}^{+0.36}) \times 10^{-4}$  &       3750--4000 & $(5.78_{-1.78}^{+2.04}) \times 10^{-5}$ \\
      4000--4250 & $(1.75_{-0.41}^{+0.50}) \times 10^{-4}$  &       4000--4300 & $(8.53_{-2.43}^{+2.61}) \times 10^{-5}$  &       4000--4300 & $(3.76_{-1.26}^{+1.43}) \times 10^{-5}$ \\
      4250--4500 & $(1.28_{-0.35}^{+0.39}) \times 10^{-4}$  &       4300--4600 & $(5.53_{-1.75}^{+1.97}) \times 10^{-5}$  &       4300--4600 & $(2.29_{-1.04}^{+0.97}) \times 10^{-5}$ \\
      4500--4750 & $(8.91_{-2.73}^{+3.19}) \times 10^{-5}$  &       4600--4900 & $(3.59_{-1.53}^{+1.47}) \times 10^{-5}$  &       4600--4900 & $(1.35_{-0.83}^{+0.64}) \times 10^{-5}$ \\
      4750--5000 & $(6.43_{-2.18}^{+2.59}) \times 10^{-5}$  &       4900--5300 & $(2.10_{-1.27}^{+1.02}) \times 10^{-5}$  &       4900--5300 & $(7.22_{-6.08}^{+3.87}) \times 10^{-6}$ \\
      5000--5300 & $(4.05_{-1.65}^{+1.91}) \times 10^{-5}$  &       5300--5800 & $(1.10_{-0.95}^{+0.64}) \times 10^{-5}$  &       5300--5800 & $(3.38_{-3.38}^{+2.10}) \times 10^{-6}$ \\
      5300--5600 & $(2.63_{-1.22}^{+1.44}) \times 10^{-5}$  &       5800--6500 & $(4.89_{-4.89}^{+3.53}) \times 10^{-6}$  &       5800--6500 & $(1.37_{-1.37}^{+1.02}) \times 10^{-6}$ \\
      5600--6000 & $(1.51_{-1.09}^{+1.00}) \times 10^{-5}$  &  &  &  & \\
      6000--6500 & $(7.76_{-7.76}^{+6.38}) \times 10^{-6}$  &  &  &  & \\
      &  &  &  &  &\\
      \hline
    \end{tabular}
    \caption{Inclusive differential neutron production cross section for p-p collisions at $\sqrt{s} = 13$~TeV, relative to three of the six regions used in the analysis. These regions are located on the large tower of the LHCf Arm2 detector. Upper and lower uncertainties, expressed as the quadratic sum of statistical and systematics contributions, are also reported.}
    \label{tab:results_large}
  \end{center}
\end{table}

\end{document}